\documentclass[11pt, a4paper]{article}

\pdfoutput=1

\usepackage{jcappub}
\usepackage{xfrac}
\usepackage{times}
\usepackage{etoolbox}
\usepackage{mathtools}
\usepackage{bm}
\usepackage{marginnote}
\renewcommand{\shortparallel}{{\mkern3mu\vphantom{\perp}\vrule depth 0pt\mkern2mu\vrule depth 0pt\mkern3mu}}
\newcommand{\D}{{\mathrm{d}}}
\newcommand{\mm}[1]{\mathrm{#1}}

\newcommand{\lw}{l_{\mm{t}}}
\newcommand{\lo}{l_\mm{osc}}
\newcommand{\blangle}{\bigl\langle}
\newcommand{\brangle}{\bigr\rangle}
\newcommand{\tr}{{\rm tr}}
\renewcommand{\vec}[1]{\mathbf{#1}}
\renewcommand{\Re}{\mbox{Re}}
\renewcommand{\Im}{\mbox{Im}}

\makeatletter
\patchcmd{\maketitle}{\@fpheader}{}{}{}
\makeatother


\arxivnumber{1611.04526}

\title{Extragalactic photon-ALP conversion at CTA energies}

\author{A. Kartavtsev$^a$,}
\emailAdd{alexander.kartavtsev@mpp.mpg.de}

\author{G. Raffelt$^a$,}
\emailAdd{raffelt@mpp.mpg.de}

\author{and H. Vogel$^{a,b}$}
\emailAdd{hvogel@slac.stanford.edu}

\affiliation[a]{Max-Planck-Institut f\"ur Physik, F\"ohringer Ring 6, 80805 M\"unchen, Germany}
\affiliation[b]{SLAC National Accelerator Laboratory, 2575 Sand Hill Road, Menlo Park, CA 94025, U.S.A.}

\abstract{ Magnetic fields in extragalactic space between galaxy clusters may induce conversions between photons and axion-like particles (ALPs), thereby shielding the photons from absorption on the extragalactic background light. For TeV gamma rays, the oscillation length ($\lo$) of the photon-ALP system  becomes inevitably of the same order as the coherence length of the magnetic field $l$ and the length over which the field changes significantly (transition length $\lw$) due to refraction on background photons. We derive exact statistical evolution equations for the mean and variance of the photon and ALP transfer functions in the non-adiabatic regime ($\lo\sim l\gg \lw$). We also make analytical predictions for the transfer functions in the quasi-adiabatic regime ($\lo\ll l, \lw$). Our results are important in light of the upcoming Cherenkov Telescope Array (CTA), and may also be applied to models with non-zero ALP masses.
}

\begin{document}

\begin{minipage}{\textwidth}
  \flushright
  MPP-2016-324\\
  SLAC-PUB-16842
\end{minipage}
\vskip1.25cm
\maketitle
\flushbottom
\newpage

\section{\label{sec:intro}Introduction}

High energy ($>100\,  \mm{GeV}$) photons are continuously emitted by blazars. During their propagation these gamma rays face an opaque wall that consists of the extragalactic background light (EBL) and leads to efficient absorption through the $\gamma\gamma^{\rm bkg}\rightarrow e^+e^-$ pair creation process~\cite{Gould:1967zza,Fazio:1970pr,Dwek:2012nb}.
The absorption rate depends on the density and energy spectrum of the EBL, which can be estimated by several methods. Whereas a direct measurement is complicated because of foreground emission~\cite{Hauser:2001xs,Dwek:2012nb}, a lower limit can be inferred by counting EBL sources~\cite{Madau:1999yh,Fazio:2004kx,Kneiske:2010pt,Dominguez:2010bv}. The inferred density of the EBL can be contrasted with an indirect measurement of the EBL by observing blazar spectra with Cherenkov telescopes like H.E.S.S., MAGIC, and VERITAS,
or with Fermi LAT. All these approaches have converged to similar results for the density
of the EBL~\cite{Aharonian:2005gh,Mazin:2007pn,Aliu:2008ay,Stecker:2008fp,Pettinari:2010ay,Belikov:2010ma,Ackermann:2012sza,Abramowski:2012ry,Abramowski:2013oea,Sinha:2014lfa,Dominguez:2015ama,Biteau:2015xpa}.
However, several authors~\cite{Protheroe:2000hp,DeAngelis:2007dqd,DeAngelis:2011id,Horns:2012fx,Meyer:2012sb,Horns:2013pha,Rubtsov:2014uga,Horns:2016vfv} have found an interesting indication that the density of the EBL as inferred from blazar spectra is below the lower limits from galaxy counts, thereby making the Universe
unexpectedly transparent to gamma rays.

This discrepancy has been tentatively interpreted as a  manifestation of  axion-like particles (ALPs)~\cite{DeAngelis:2007dqd,DeAngelis:2008sk,SanchezConde:2009wu,Dominguez:2011xy,Galanti:2015rda}, although other suggestions exist~\cite{Essey:2009zg,Zheng:2013lza,Zheng:2015gfw,Dzhatdoev:2016ftt}. Axion-like particles are hypothetical pseudo-scalar particles that couple to two photons, which permits photons and ALPs to interconvert in a magnetic field background. Because ALPs are not absorbed by the EBL, the photons that oscillate into ALPs are protected from absorption, and can be converted back into photons close to the source. This mechanism reduces the effective absorption strength experienced by the photons in the intergalactic space. It is similar to the photon regeneration technique (``light shining through a wall''), e.g.\ the ALPS Experiment at DESY~\cite{Bahre:2013ywa}, with the EBL taking the role of the opaque wall that separates the source from the detector.

The efficiency of the photon-ALP conversion depends on the strength of the external  magnetic field  as well as the
time the photon-ALP system is exposed to its influence. The terrestrial experiments usually feature environments with
well controlled magnetic fields that spread over several meters and have strength of the order of a few Tesla. In contrast to these well controlled conditions, photons from blazars experience much more variable environments.  Photon-ALP conversion may occur in the magnetic fields of the source or the host galaxy and cluster~\cite{Hochmuth:2007hk,Simet:2007sa,Hooper:2007bq,DeAngelis:2007wiw,Chelouche:2008ta,Chelouche:2008ax,Jimenez:2011pg,Horns:2012kw,Wouters:2012qd,Galanti:2013afa,Brun:2013vfa,Wouters:2013eka,Meyer:2014epa,SanchezConde:2009wu} as well as in the tiny ($\lesssim \, \mm{nG}$~\cite{DeAngelis:2007rw,Pshirkov:2015tua}) magnetic fields in extragalactic space~\cite{DeAngelis:2007dqd,DeAngelis:2008sk,SanchezConde:2009wu,Dominguez:2011xy,Galanti:2015rda}, which may be coherent over very large (Mpc) scales~\cite{Kronberg:1993vk,Grasso:2000wj,Furlanetto:2001gx,Dolag:2004kp,Bertone:2006mr,Widrow:2011hs,Durrer:2013pga}.

For photon-ALP propagation in extragalactic space  predictions are made complicated by the fact that we only have limited knowledge about the magnetic field along the line of sight to gamma ray sources.
For this reason, the photon and ALP transfer functions have been computed  in a statistical sense by simulating many
different magnetic field configurations along the line of sight and computing the resulting mean and variance of the
gamma ray  flux, see e.g.\ references~\cite{Mortsell:2002dd,Mirizzi:2009}. Usually, additional assumptions about the distribution of the extragalactic magnetic field are made. In particular, a domain-like
structure was assumed in references~\cite{DeAngelis:2007dqd,DeAngelis:2011id,DeAngelis:2008sk,Mirizzi:2009,Bassan:2010ya}, i.e.\ each domain was taken to have a fixed length $l$ equal to the magnetic field's coherence
length, and that within each domain the magnetic field is constant. At the border between the domains the magnetic field was
modelled to change its direction discontinuously while having a fixed absolute value. With this set of approximations, Mirizzi and Montanino~\cite{Mirizzi:2009} derived a system of differential equations for the mean and variances that reproduces the results of the computationally demanding Monte-Carlo simulations.

\begin{figure}[t!]
  \center
  \includegraphics[width=0.45\textwidth]{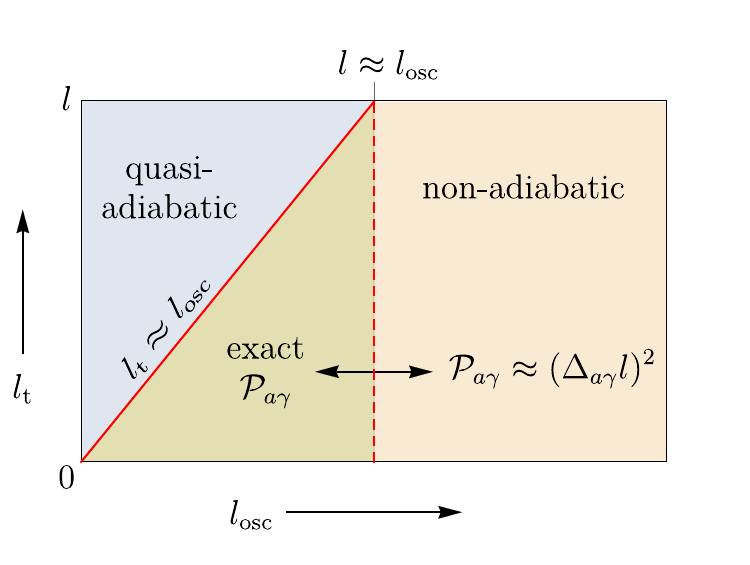}
  \caption{\label{fig:PhaseDiag}Qualitatively different regimes of photon-ALP propagation. On the horizontal axis we plot the oscillation length increasing to the right. The vertical axis shows the transition width between two magnetic field domains $\lw$. It is bounded from above by the domain size $l$. The red solid line indicates the transition between quasi-adiabatic and non-adiabatic propagation. It is given by $\lw \approx \lo$. In the non-adiabatic regime the result of Mirizzi and Montanino~\cite{Mirizzi:2009} is valid for  $l\ll \lo$, which is fulfilled to the right of the red dashed line given by $l\approx \lo$. This region is denoted as ``$\mathcal{P}_{a\gamma}\approx (\Delta_{a\gamma}l)^2$''. The region denoted as ``exact $\mathcal{P}_{a\gamma}$'' requires an extension of the ansatz of reference~\cite{Mirizzi:2009} which is derived in section~\ref{sec:averaged} of the present work.}
\end{figure}

The approximations used in these works~\cite{DeAngelis:2007dqd,DeAngelis:2011id,DeAngelis:2008sk,Mirizzi:2009,Bassan:2010ya} were believed to be justified because of the following reasoning. First, at the time when e.g.\ reference~\cite{Mirizzi:2009} was published, one expected that at TeV energies the diagonal elements of the photon-ALP Hamiltonian become small compared to its off-diagonal elements, $\Delta_{a\gamma}\sim g_{a\gamma}B$, with photon-ALP coupling constant $g_{a\gamma}$ and magnetic field strength $B$. In other words, the photon-ALP mixing was expected to be close to maximal at TeV energies, with the oscillation probability per domain simplifying to $\mathcal{P}_{a\gamma}\approx (\Delta_{a\gamma} l)^2$.
Second, for the typical strength of the extragalactic magnetic fields and values of the photon-ALP coupling consistent with the
existing constraints~\cite{Dias:2014osa}, the resulting oscillation length  $\lo \approx 200 \, \mm{Mpc}$  was expected to be much larger than the typical domain size $l$ and the transition length $\lw$ between two domains. Thus,
even a smooth transition from one domain to another would be perceived as abrupt and non-adiabatic by the photon-ALP system,
which justified the approximation of discontinuous transitions between the domains. The range of parameters for which
this approximation is applicable is shown on the right-hand side of the red dashed line in figure~\ref{fig:PhaseDiag}.

The other two qualitatively distinct parameter regions in figure~\ref{fig:PhaseDiag}, labeled as ``exact $\mathcal{P}_{a\gamma}$'' and ``quasi-adiabatic''
respectively, are absent in the setup considered in references~\cite{DeAngelis:2007dqd,DeAngelis:2011id,DeAngelis:2008sk,Mirizzi:2009,Bassan:2010ya}. Even if the ALP mass is negligible, these regions appear due to an additional contribution
to the diagonal elements of the photon-ALP Hamiltonian that stems from the recently identified forward scattering on background photons~\cite{Dobrynina:2014qba}.
This additional term grows linearly with the energy of the gamma ray and, for typical extragalactic magnetic fields and TeV energies, becomes larger than the mixing term $\Delta_{a\gamma}$, i.e.\ it is especially relevant for the energy range that the upcoming generation of Cherenkov telescopes like CTA~\cite{Consortium:2010bc}, HAWC~\cite{Abeysekara:2013tza}, and HiSCORE~\cite{Tluczykont:2011wq} will be sensitive to.
In the present work we take a closer look at the qualitative impact of this contribution on the photon-ALP oscillations in
extragalactic magnetic fields. In particular, we expand on the considerable progress in the analytical description of oscillations between photons and ALPs~\cite{Raffelt:1987im,Carlson:1994yqa,Deffayet:2001pc,Grossman:2002by,Csaki:2003ef,Lai:2006af,Ganguly:2008kh,Agarwal:2008ac,Agarwal:2009ic,Mirizzi:2009,Reesman:2014ova, Wang:2015dil} by treating absorption and dispersion rigorously. Because the additional contribution suppresses the oscillation length such that \smash{$\lo\approx 80\,\frac{\mm{TeV}}{\omega}$} Mpc for $\omega \gtrsim 3$ TeV, we also present an interpretation of the various phenomena and issues of photon-ALP oscillations that arise in the regime where $\lo$  becomes of similar size as $l$ and~$\lw$.

In the region labeled as ``exact $\mathcal{P}_{a\gamma}$'' in figure~\ref{fig:PhaseDiag}, the oscillation length fulfills $\lo\sim l \gg \lw$. Here, the propagation still proceeds non-adiabatically, so that the transition between two magnetic field domains does not have to be modeled explicitly.  In section~\ref{sec:averaged}, we show how to extend the formalism of Mirizzi and Montanino~\cite{Mirizzi:2009}, which was derived assuming $\Delta_{\mm{osc}}l\ll 1$ (equivalent to $l\ll \lo$),  by treating dispersion and absorption exactly. The resulting
equations are surprisingly simple in that the most important change is the substitution of the expanded $\mathcal{P}_{a\gamma}$
by its exact expression presented in section~\ref{sec:averaged}. Moreover, we find that the formalism is also applicable if the absolute value of the magnetic field is different in different domains, as long as the magnetic field does not have a preferred direction, and that it gives good estimates even if the domain lengths vary from domain to domain.

At high energies the oscillation length approaches the transition length between magnetic field domains. In this regime, the photon-ALP system becomes very sensitive to the exact configuration of the magnetic field along the line of sight, and one has to study photon-ALP oscillations numerically. At even higher energies,
once the oscillation length becomes much smaller than the transition length between the domains, the propagation of
the photon-ALP system becomes close to adiabatic (the region ``quasi-adiabatic'' in figure~\ref{fig:PhaseDiag}). As we
demonstrate in section~\ref{sec:quasiadiabatic}, it is possible to understand the behavior of the resulting photon transfer
function analytically. For exactly adiabatic propagation the transfer function depends only on the magnetic field
strength at the source and the detector, whereas in the presence of a small non-adiabaticity and of  strong absorption it also depends on the first derivative of the magnetic field at the emission and detection points.

In section~\ref{sec:conclusions}, we summarize these findings and present our conclusions. Additionally, we discuss possible future developments of our study.

\section{\label{sec:averaged}Statistical approach to photon-ALP propagation}

The photon and ALP transfer functions could be easily calculated if we knew the exact configuration of the magnetic field  along the line of sight to a gamma-ray source. Unfortunately, our knowledge of extragalactic magnetic fields is limited~\cite{Durrer:2013pga,Neronov:1900zz} and not even the approximate strength and coherence length are known. We have to rely on models of extragalactic magnetic field to study the impact of photon-ALP oscillations on the photon propagation. A popular model, which has also been used by De Angelis, Roncadelli, and Mansutti in reference~\cite{DeAngelis:2007dqd}, is due to Furlanetto and Loeb~\cite{Furlanetto:2001gx}. In this model the extragalactic magnetic field is generated by quasar outflows, which form bubbles with magnetic fields that typically spread over $\sim 4\, \rm{Mpc}$ with field strengths of the order of $1~\rm{nG}$.  Alternatively, extragalactic magnetic fields may originate from inflation, in which case the magnetic field power spectrum could be scale-invariant at large scales~\cite{Grasso:2000wj,Widrow:2011hs,Durrer:2013pga}. In the following, we assume that the typical coherence length of the magnetic field is $l=10\, \mm{Mpc}$.

Because these models predict statistical properties of the magnetic field, they can be used to infer the probability distribution of the photon and ALP transfer functions.
This is achieved by simulating a large number of magnetic field configurations along the line of sight and solving
the propagation equations for each of the generated  field configurations individually~\cite{Csaki:2001yk,Mortsell:2002dd,Mortsell:2003ya}. The mean and variance of the transfer functions is obtained by aggregating the data \cite{DeAngelis:2007dqd}. This approach is computationally demanding but flexible with regards to the magnetic field input.

As has been noted by Mirizzi and Montanino~\cite{Mirizzi:2009}, for specific configurations of the magnetic field, namely if it has a constant comoving domain length and the transition region between individual domains is infinitely thin (magnetic field with ``hard edges''), results of the Monte Carlo simulation for the mean are reproduced by the solution of two coupled differential equations for the photon and ALP transfer functions.
When Mirizzi and Montanino published their work, both of these assumptions were thought to be justified for gamma rays with TeV energies if the mass of the ALP was small enough. They defined a critical energy above which the photon-ALP mixing is close to maximal. In this regime the oscillation length is determined by the magnetic field and typically much smaller than the typical length scales associated with the magnetic field. This condition
was especially important for the derivation presented in reference~\cite{Mirizzi:2009}.
However, forward scattering on CMB photons recently identified in reference~\cite{Dobrynina:2014qba} ensures that the oscillation length decreases as the energy increases and prevents the mixing from becoming maximal for TeV gamma rays. In this section we demonstrate that the approach of reference~\cite{Mirizzi:2009} can nevertheless be extended to the case of arbitrary mixing angles as long as the
oscillation length remains large compared to the width of the transition region between the domains.

\paragraph{The evolution equation and Hamiltonian matrix.}

\begin{figure}[t!]
  \center
  \includegraphics[width=0.45\textwidth]{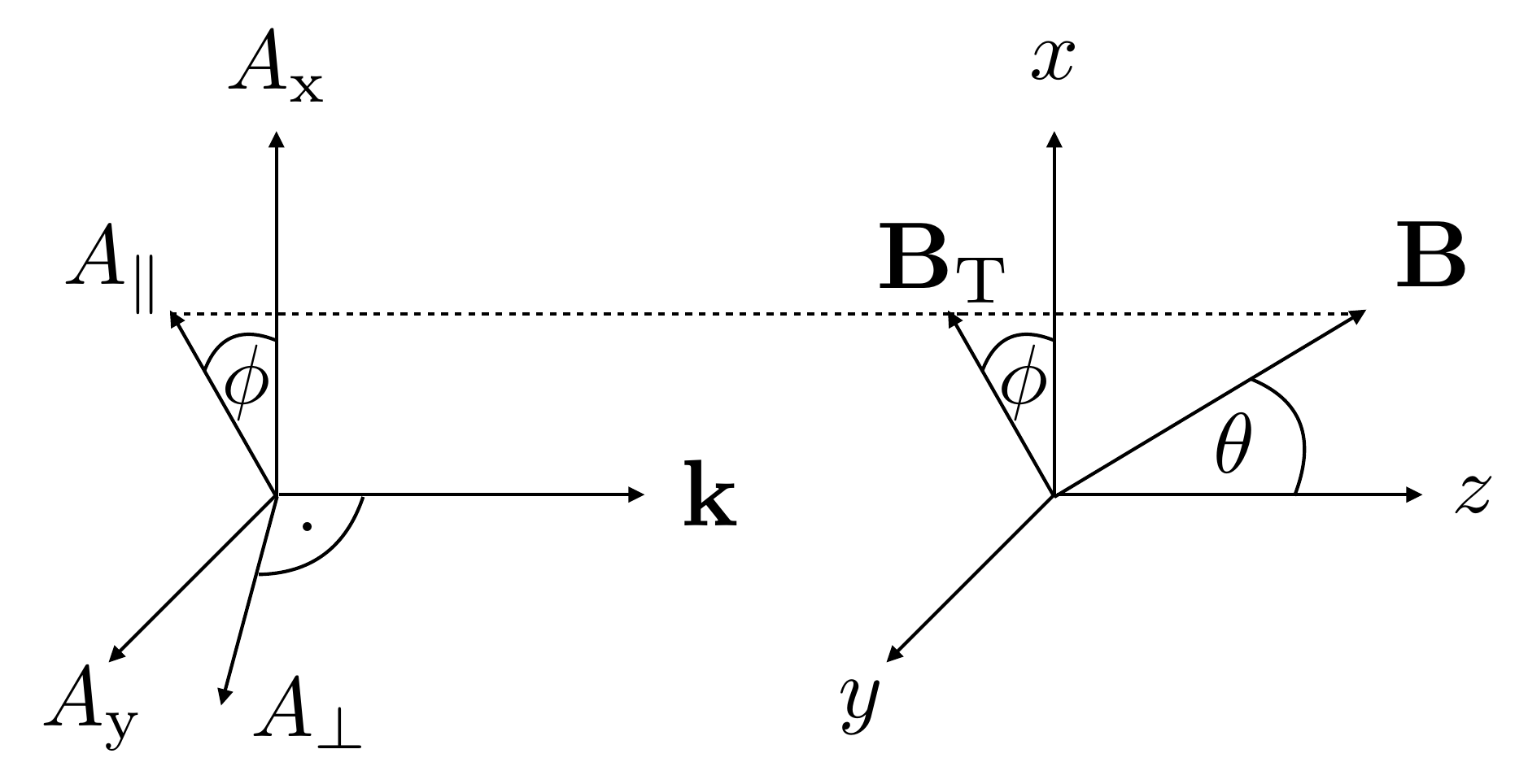}
  \caption{\label{fig:coordinate} Geometry of the photon polarizations and the magnetic field. The direction of photon propagation ${\bf{k}}$ is chosen to point along the $z$-axis. The linear polarizations of the photon and the projection of the magnetic field that is responsible for photon-ALP conversions ${\bf{B}}_\mm{T}$ lie in the $xy$-plane with polar angle $\phi$.  $\theta$ denotes the zenith angle. The linear combination of polarizations vectors $A_x$ and $A_y$ that is parallel to ${\bf{B}}_{\rm T}$ is indicated by $A_\parallel$. $A_\perp$ is the perpendicular component.}
\end{figure}

The evolution equation for a photon-ALP system propagating in the $z$-direction in a magnetic field background
reads~\cite{Raffelt:1987im,Csaki:2003ef}
\begin{align}
  \label{SchroedEq}
  i\frac{\D}{\D z}\mathbf{A}=\left(\mathbf{H}_{\rm dis}-\frac{i}{2}\mathbf{H}_{\rm abs}\right)\mathbf{A}\,.
\end{align}
Note that here and throughout the manuscript we use boldface to denote vectors and matrices.
In equation~\eqref{SchroedEq} the three-component wave function,
\begin{align}
  \mathbf{A}\equiv
  \begin{pmatrix}
    A_\perp \\
    A_\shortparallel \\
    a
  \end{pmatrix},
\end{align}
contains the two states of linear polarization perpendicular and parallel to the external field,
$A_\perp$ and $A_\shortparallel$, as well as the ALP amplitude, $a$. For clarification we present the geometry of the photon polarizations and the magnetic field in figure~\ref{fig:coordinate}.

The high-energy photons are absorbed in the process $\gamma\gamma^{\rm bkg}\rightarrow e^+e^-$. In an (approximately) isotropic and unpolarized  EBL both photon polarizations are absorbed with the same strength $\Gamma$. The ALPs can in principle also be directly absorbed in collisions with
the extragalactic medium. However, the resulting absorption rate is quadratic in the photon-ALP coupling
and is negligibly small. Thus the absorptive part of the Hamiltonian can be written
in the form
\begin{align}
  \label{Habs}
  \mathbf{H}_{\rm abs}=
  \begin{pmatrix}
    \Gamma & 0 & 0\\
    0 & \Gamma & 0 \\
    0 & 0 & 0
  \end{pmatrix}.
\end{align}
To leading order in the fine-structure constant the photon absorption rate is given by~\cite{Mirizzi:2009}
\begin{subequations}
  \begin{align}
    \Gamma(\omega)&=\int^\infty_{m^2_e/\omega} \D\epsilon\, \frac{\D n^{\rm bkg}_\gamma}{\D\epsilon}
    \int^{1-\frac{2m^2_e}{\omega\epsilon}}_{-1}\D\xi\, \frac{1-\xi}{2}\sigma_{\gamma\gamma}(\beta)\,,\\
    \sigma_{\gamma\gamma}(\beta)&=\sigma_0(1-\beta^2)
    \left[2\beta(\beta^2-2)+(3-\beta^4)\log\frac{1+\beta}{1-\beta}\right]\,,
  \end{align}
\end{subequations}
where $\sigma_0=3\sigma_\mm{T}/16\approx1.3\times 10^{-25}$ cm$^2$~\cite{Franceschini:2008tp} with the Thomson cross section $\sigma_\mm{T}$, $\epsilon$ is the energy of the background photon, $\xi$ is the cosine of the angle between the incident and the background
photons, and $\beta=[1-4m^2_e/s]^\frac12$ with $s=2\omega\epsilon(1-\xi)$ is the electron velocity in
the center of mass frame.
To obtain the optical depth, the absorption rate has to be integrated over the
distance or, equivalently, redshift. To date, blazars are observed at redshifts of maximally
$z\approx 0.944$~\cite{Sitarek:2015jza} for which the redshift dependence of energies, densities and magnetic fields have to be taken into account. Nevertheless, we will neglect redshift in the following because our goal is a conceptual one that would be merely clouded by this additional layer of complication. For our purposes, the optical depth is just the average absorption rate times the distance to the source.
For crude numerical estimates at zero redshift we use for the absorption
rate~\cite{Mirizzi:2009}
\begin{align}
  \label{GammaApprox}
  \frac{\Gamma(\omega)}{\mathrm{Mpc}^{-1}}\simeq 1.1 \times 10^{-3}
  \left(\frac{\omega}{\rm TeV}\right)^{1.55}\,.
\end{align}
Although the underlying EBL model might now be disfavored~\cite{Aliu:2008ay}, this analytic expression suffices for our qualitative analysis.

Choosing the coordinate system
such that the magnetic field $\vec B$ lies in the $xz$-plane, see figure~\ref{fig:coordinate},  we arrive at~\cite{Raffelt:1987im,Mirizzi:2009}
\begin{align}
  \label{Hdis}
  \mathbf{H}_{\rm dis}=
  \begin{pmatrix}
    \Delta_\perp & 0 & 0 \\
    0 & \Delta_\shortparallel &  \Delta_{a\gamma} \\
    0 & \Delta_{a\gamma} & \Delta_a
  \end{pmatrix}\,,
\end{align}
where we have neglected the tiny contribution from Faraday rotation. Both $\Delta_\perp$
and $\Delta_\shortparallel$ receive contributions from refraction on the electron plasma,
refraction on the magnetic field (which can be viewed as forward scattering on virtual photons) and, as has been noted in
reference~\cite{Dobrynina:2014qba}, refraction on real photons (e.g.\ CMB and EBL),
\begin{subequations}
  \begin{align}
    \Delta_\perp&=\Delta_{\rm pl}+2\Delta_{{\rm B}}+\Delta_{\gamma\gamma}\,,\\
    \label{DeltaParal}
    \Delta_\shortparallel&=\Delta_{\rm pl}+\tfrac72\Delta_{\rm B}+\Delta_{\gamma\gamma}\,.
  \end{align}
\end{subequations}
The first two contributions read~\cite{Mirizzi:2009},
\begin{subequations}\label{eq:deltaold}
  \begin{align}
    \Delta_{\rm pl}&=-\frac{2\pi\alpha n_e}{m_e} \omega^{-1} \approx\,-1.1\times10^{-11} \left(\frac{\omega}{{\mm{TeV}}}\right)^{-1} \left(\frac{n_e}{{10^{-7} \, {\mm{cm}}^{-3}}}\right)\,{\mm{Mpc}}^{-1}, \\
    \Delta_{\rm B}&=\frac{24\alpha^2}{135}\frac{\rho_{\rm B}}{m^4_e}\sin^2\theta\, \omega\approx 4.1\times 10^{-9} \sin^2\theta\left(\frac{\omega}{{\mm{TeV}}}\right)\left( \frac{B}{{\mm{nG}}}\right)^2\, {\mm{Mpc}}^{-1}\label{eq:deltaB}\,,
  \end{align}
\end{subequations}
where $n_e$ is the electron density, $\rho_{\rm B}\equiv \frac12 \vec{B}^2$ is the energy density of the magnetic field
and $\theta$ is the polar angle of $\vec{B}$. The index of refraction induced by an electromagnetic field (forward scattering on virtual photons) is proportional to its energy density~\cite{Tarrach:1983cb,Barton:1989dq,Latorre:1994cv,Kong:1998ic,Thoma:2000fd}, see equation~\eqref{eq:deltaB}. Similarly, the contribution of forward scattering on real photons is proportional to the energy density of the background
photons~\cite{Dobrynina:2014qba}  and is therefore dominated by the CMB, whose energy density
is roughly an order of magnitude larger than that of the EBL,
\begin{align}
  \label{DeltaGamma}
  \Delta_{\gamma\gamma}\approx \frac{44\alpha^2}{135}\frac{\rho_{\rm CMB}}{m^4_e} \omega
  \approx 8.0\times 10^{-2} \left(\frac{\omega}{{\mm{TeV}}}\right)\, {\mm{Mpc}}^{-1}  \,.
\end{align}
The remaining diagonal element of the dispersive Hamiltonian, $\Delta_a=-m_a^2/2\omega$, is due to
the ALP mass. From the energy dependence of $\Delta_a$ it follows that there is an energy from which on $\Delta_{\gamma\gamma}$ dominates the trace of the dispersive Hamiltonian matrix. In the following, we will always neglect any contribution from $\Delta_a$ whenever we show numerical results, i.e.\ we assume that the ALP mass is effectively zero. This does not affect the analytical results shown below.

The off-diagonal elements that couple ALPs to photons are proportional to the photon-ALP
coupling constant and strength of the background magnetic field,
\begin{align}\label{eq:deltaag}
  \Delta_{a\gamma}=\frac{g_{a\gamma}}{2}B_\mm{T}\approx 1.5\times 10^{-2} \sin\theta \left(\frac{g_{a\gamma}}{10^{-11} \, {\mm{GeV}}^{-1}}\right)\left(\frac{B}{{\mm{nG}}}\right)\, {\mm{Mpc}}^{-1}\,,
\end{align}
where $B_\mm{T}$ is the projection of the magnetic field on the $xy$-plane.
This mixing term is decisive for photon-ALP conversion. At the same time it
is the one that is subject to the largest uncertainty, as very little is known about extragalactic
magnetic fields and the photon-ALP coupling. Comparing the numerical values of the diagonals and the off-diagonals of the Hamiltonian matrix, we conclude that the assumptions that the diagonals of $\bm{H}_{\rm dis}$ can be neglected ceases to be valid at high energies, and the formalism developed in reference~\cite{Mirizzi:2009} has to be generalized.

\paragraph{Transfer function within a single domain.}

Following reference~\cite{Mirizzi:2009} we introduce
the density matrix  $\bm{\rho}=\mathbf{A}\otimes\mathbf{A}^\dagger$. Its diagonal elements are the usual
number densities of the two photon polarization states and of the ALP respectively, while the
off-diagonal terms contain information on the coherence between these states~\cite{Sigl:1992fn}.
As follows from equation~\eqref{SchroedEq}, its evolution is described by
\begin{align}
  \label{EquationForRho}
  i\frac{\D}{\D z}\bm{\rho}=\mathbf{H}\bm{\rho}-\bm{\rho}\mathbf{H}^\dagger=
  \bigl[\mathbf{H}_{\rm dis},\bm{\rho}\bigr]-\frac{i}{2}\bigl\{\mathbf{H}_{\rm abs},\bm{\rho}\bigr\}\,,
\end{align}
where, because of the photon absorption, we have to distinguish between the Hamiltonian and its
conjugate.

Within each domain the magnetic field and the Hamiltonian matrix are assumed to be constant. This approximation is
well motivated as long as the photon-ALP oscillation length remains much larger than the typical domain size. In this case the evolution
equation~\eqref{SchroedEq} can be solved analytically without any further approximations,
\begin{align}
  \label{ExactSol}
  \mathbf{A}(z)=\mathbf{U}(z)\mathbf{A}(0) \quad {\rm with} \quad
  \mathbf{U}(z)=\exp(-i\, \mathbf{H} z)= \mathbf{O}\exp(-i\, \mathbf{E} z)\,\mathbf{O}^T\,,
\end{align}
where
\begin{align}
  \label{ComplexOrthogonalTrafo}
  \mathbf{O}=
  \begin{pmatrix}
    1 & 0 & 0\\
    0 & c_\alpha & s_\alpha \\
    0 & -s_\alpha & c_\alpha
  \end{pmatrix}
\end{align}
is a complex orthogonal transformation that diagonalizes the Hamiltonian matrix. Here and in the following we use the notation $c_\alpha \equiv \cos \alpha$ and $s_\alpha\equiv \sin \alpha$. The (complex) photon-ALP
mixing angle $\alpha$ is determined by
\begin{align}
  \label{MixingAngle}
  \tan 2\alpha=\frac{2\Delta_{a\gamma}}{\Delta_\shortparallel-\Delta_a-\frac{i}{2}\Gamma}\,.
\end{align}
Note that because $\Delta_a < 0$ and $\Delta_\shortparallel > 0$ the contributions of the ALP mass and that
of the photon-photon refraction always add up. They cannot cancel each other and must be separately small to
achieve large mixing. If this is the case and $\Delta_{a\gamma}\gg \Delta_\shortparallel-\Delta_a$, then the
photon-ALP mixing is close to maximal, $\alpha\rightarrow \pi/4$ (if the absorption is small as well). On the other hand, because $\Delta_{\gamma\gamma}$
grows linearly with the photon energy, see equation~\eqref{DeltaGamma}, at sufficiently high energies $\Delta_{a\gamma}\ll \Delta_\shortparallel-\Delta_a$
and the photon-ALP mixing is small.

The diagonalization leaves the $(1,1)$ element of the Hamiltonian invariant and therefore the $(1,1)$
element of $\mathbf{E}={\rm diag}(E_1,E_2,E_3)$ is also given by
$E_1=\Delta_\perp -\frac{i}{2}\Gamma$. The $(2,2)$ and $(3,3)$ elements are given by
$E_{2,3}=\frac12\left(\Delta_\shortparallel+\Delta_a\pm \Delta_\mm{osc}-\tfrac{i}2\Gamma\right)$, where
\begin{align}
  \label{DeltaDef}
  \Delta_\mm{osc}=\bigl[(\Delta_\shortparallel-\Delta_a-\tfrac{i}2\Gamma)^2+(2\Delta_{a\gamma})^2\bigr]^\frac12\,
\end{align}
is  the generalized (in that it includes absorption) photon-ALP oscillation frequency. For TeV gamma rays it is dominated by $\Delta_{\gamma\gamma}$
and grows linearly with energy. The oscillation length, which is inversely proportional to $\Delta_\mm{osc}$,
therefore becomes smaller at high energies.

Performing the matrix multiplication in equation~\eqref{ExactSol} we arrive at
\begin{align}
  \label{EvolutionOperator}
  \mathbf{U}(z)=
  \begin{pmatrix}
    e^{-iE_1z} & 0 & 0 \\
    0 & c^2_\alpha e^{-iE_2z} + s^2_\alpha e^{-iE_3z} & c_\alpha s_\alpha \bigl(e^{-iE_2z}-e^{-iE_3z}\bigr)  \\
    0 &  c_\alpha s_\alpha \bigl(e^{-iE_2z}-e^{-iE_3z}\bigr)  & s^2_\alpha e^{-iE_2z} + c^2_\alpha e^{-iE_3z} \\
  \end{pmatrix}\,.
\end{align}
In the language of reference~\cite{Mirizzi:2009}, where the solution was obtained by expanding the exponent in
equation~\eqref{ExactSol} to the second order in the Hamiltonian, equation~\eqref{EvolutionOperator} resums all
orders in dispersion as well as in absorption. The corresponding solution for the density matrix is obtained
by multiplying equation~\eqref{ExactSol} by its Hermitian conjugate,
\begin{align}
  \label{ExactSolForRho}
  \bm{\rho}(z)=\mathbf{U}(z)\bm{\rho}(0)\mathbf{U}^\dagger(z)\,.
\end{align}
Because the states of linear polarization parallel and orthogonal to the external field
interact differently, see equation~\eqref{Hdis}, the polarization of the final photon non-trivially depends on the relative angle $\phi$ of the projection of the magnetic field orthogonal to the direction of motion and the initial polarization of the incoming photons. We usually have limited information about $\phi$. For a fixed magnetic field direction and an unpolarized source, we may still compute an expectation value for the number of photons after trespassing the domain,
\begin{align}
  \label{ExactSolForRhoAv}
  \langle\bm{\rho}(z)\rangle_{\phi}=\langle\mathbf{U}(z)\bm{\rho}(0)\mathbf{U}^\dagger(z)\rangle_{\phi}\,,
\end{align}
by performing a statistical average over $\phi$ that appears in the initial condition via
\begin{align}
  \label{InitialState}
  \mathbf{A}(0)\sim
  \begin{pmatrix}
    s_\phi c_\beta\\
    c_\phi c_\beta\\
    s_\beta
  \end{pmatrix}\,.
\end{align}
Here, $\beta$ parametrizes the ALP admixture. The elements of $\mathbf{A}(0)$ contain in
principle complex phases  but these are irrelevant for the argument and are not indicated for the
sake of brevity. Averaging $\rho(0)$ over $\phi$ we obtain
\begin{align}
  \langle \bm{\rho}(0)\rangle_\phi = \frac{1}{2\pi}\int_0^{2\pi} \! \! \! \D\phi\, \bm{\rho}(0) =
  {\rm diag}\big[\textstyle{\frac12}T_\gamma(0),\textstyle{\frac12}T_\gamma(0),T_a(0)\bigr]\,,
\end{align}
where we introduced photon and ALP transfer functions, $T_\gamma\equiv
\langle \rho_{11}\rangle + \langle \rho_{22}\rangle$ and $T_a\equiv\langle \rho_{33}\rangle$. This definition is justified  because the different photon polarizations of gamma rays are not distinguished in current experiments.
Substituting these expressions into equation~\eqref{ExactSolForRhoAv} we arrive at
\begin{subequations}
  \label{SolForTransferFunctions}
  \begin{align}
    \label{SolForTg}
    T_\gamma(z)-T_\gamma(0)&=-\mathcal{P}_{a\gamma}(z)\bigl[\textstyle{\frac12}\,T_\gamma(0)-T_a(0)\bigr]
    -\textstyle{\frac12} \bigl[1-e^{-\Gamma z}+\delta_\gamma(z)\bigr]T_\gamma(0)
    \\
    \label{SolForTa}
    T_a(z)-T_a(0)&=+\mathcal{P}_{a\gamma}(z)\bigl[\textstyle{\frac12}\,T_\gamma(0)-T_a(0)\bigr]-
    \delta_a(z)T_a(0)\,,
  \end{align}
\end{subequations}
which is a closed system of equations. In equation~\eqref{SolForTransferFunctions}
\marginnote{\scriptsize Typo correction:\\ $\text{1/2}$ in exponential.\\
Thanks to\\ Deep Jyoti Das.}
\begin{align}
  \label{OscProb}
  \mathcal{P}_{a\gamma}(z) = |U_{23}(z)|^2 = {e^{-\Gamma z/2}}
  \bigl|\sin (2\alpha) \sin\left(\Delta_\mm{osc} z/2\right)\bigr|^2
\end{align}
is the generalized (in that it takes into account absorption) photon-ALP oscillation
probability, and
\begin{subequations}
  \begin{align}
    \delta_\gamma(z) & = 1 -  |U_{22}(z)|^2 -  |U_{23}(z)|^2\,,\\
    \delta_a(z) & = 1 -  |U_{33}(z)|^2 -  |U_{23}(z)|^2\,,
  \end{align}
\end{subequations}
where we have used $U_{32}=U_{23}$. In the absence of absorption the evolution operator, equation~\eqref{EvolutionOperator},
is unitary and therefore $\sum_i |U_{ij}|^2=\sum_j |U_{ij}|^2=1$. This implies that $\delta_\gamma, \delta_a\rightarrow 0$ for $\Gamma \rightarrow 0$.
This is consistent with the trace of equation~\eqref{EquationForRho},
\begin{align}
  \label{EqForTrace}
  \frac{\D}{\D z}(T_\gamma+T_a)=\frac{\D}{\D z}\tr\rho=-\Gamma T_\gamma\,,
\end{align}
whose right-hand side vanishes in the limit $\Gamma\rightarrow 0$.

\begin{figure}[t!]
  \center
  \includegraphics[width=0.45\textwidth]{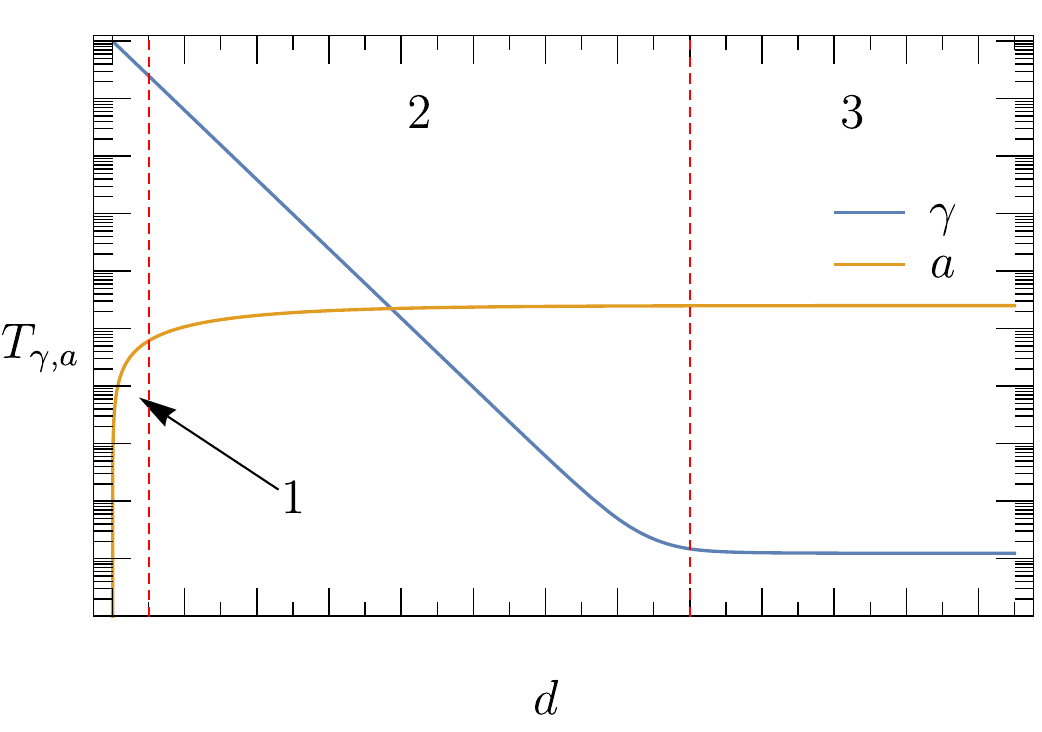}
  \caption{\label{fig:categories}Typical photon (solid blue line) and ALP (solid orange line) transfer functions as a function of the
    distance from the source. The initial condition consists of an unpolarized photon state without an ALP admixture.
    We identify three propagation regimes separated by red dashed lines: (1) photon-dominated regime, (2) intermediate regime,
    and (3) ALP-dominated regime. The transition from one regime to another is marked by a change of the slope of
  one of the transfer functions.}
\end{figure}

Alternatively, and this lies in the core of the approach developed in reference~\cite{Mirizzi:2009}, the averaging  over $\phi$ that led to equation~\eqref{SolForTransferFunctions} may be interpreted in a
different way. We fix the polarization of an incoming photon and average over the orientation of the magnetic field orthogonal to the photon's direction of motion. Equation~\eqref{SolForTransferFunctions} follows if the statistical distribution of the magnetic field does not have a preferred direction. We may also average over the remaining degrees of freedom of the magnetic field by substituting $\mathcal{P}_{a\gamma}(z)\rightarrow\langle \mathcal{P}_{a\gamma}(z)\rangle_{\bf{B}}$, and similarly for $\delta_\gamma$ and $\delta_a$, where the subscript $\bf{B}$ denotes the statistical average over all magnetic field components. The obtained equation generalizes the result of reference~\cite{Mirizzi:2009}
for propagation over a single domain. While the approach of reference~\cite{Mirizzi:2009} only includes terms that are linear and quadratic in the Hamiltonian matrix, equation~\eqref{SolForTransferFunctions} resums all orders in dispersion and absorption.

\paragraph{Average transfer functions after crossing several domains.}

The photon transfer function after crossing  one domain of length $l$ is found by setting $z=l$ in equation~\eqref{SolForTg}. If the distribution of the magnetic field in adjacent domains is statistically independent, then in order to obtain the transfer functions after traversing several domains, we may  iterate equation~\eqref{SolForTransferFunctions} using statistically averaged coefficients. Numerical results obtained in this way are in excellent agreement with results of direct Monte Carlo simulations.

The typical  dependence of the photon and ALP transfer functions on the distance from the source is presented in figure~\ref{fig:categories}.
Initially, in the photon-dominated regime, the population of photons (solid blue line) is larger than that of ALPs (solid orange line). The photons are absorbed in collisions with the EBL  and simultaneously oscillate into ALPs, thus producing a nonzero ALP population. After a short period the ALP abundance saturates and the system enters the intermediate regime. In this regime the evolution of photons is almost independent of the evolution of ALPs and vice versa. On the one hand, the number of photons converted into ALPs is small compared to the already produced ALP population. On the other hand, the number of photons created by back-conversion of the ALPs is negligibly small compared to their total number. As the photon population further decays, the photons created by back-conversion of the ALPs begin to dominate the total photon abundance. The resulting change of the slope of the photon transfer function marks the onset of the ALP-dominated regime.

\paragraph{Differential equation for transfer functions.} Instead of iterating equati\-on~\eqref{SolForTransferFunctions}, we may follow reference~\cite{Mirizzi:2009} and convert equation~\eqref{SolForTransferFunctions} into a system of coupled differential equations by approximating $T(z+l)-T(z)\approx l\,\partial_z T(z)$. This yields
\begin{subequations}
  \label{StatAveragedEqs}
  \begin{align}
    \label{StatAveragedEqsTg}
    l \frac{\D}{\D z} T_\gamma(z)&=-\bigl\langle\mathcal{P}_{a\gamma}(l)\bigr\rangle
    \bigl[\tfrac12\,T_\gamma(z)-T_a(z)\bigr]
    -\tfrac12 \bigl\langle 1-e^{-\Gamma l}+\delta_\gamma(l) \bigr\rangle T_\gamma(z)\,,\\
    \label{StatAveragedEqsTa}
    l \frac{\D}{\D z}T_a(z)&=+\bigl\langle\mathcal{P}_{a\gamma}(l)\bigr\rangle \bigl[\tfrac12
    \,T_\gamma(z)-T_a(z)\bigr]- \bigl\langle\delta_a(l)\bigr\rangle T_a(z)\,,
  \end{align}
\end{subequations}
where the averaging $\langle \bullet \rangle$ is over the direction and strength of the magnetic field.
Equation~\eqref{StatAveragedEqs} reduces to the result of reference~\cite{Mirizzi:2009} for $\Delta_\mm{osc}\approx \Delta_{a\gamma}$ and $\Gamma l\ll 1$, with the transition probability in one domain simplifying to
$\mathcal{P}_{a\gamma}(l) \approx \left(\Delta_{a\gamma}l\right)^2$ in this limit.

Equation~\eqref{StatAveragedEqs}  predicts a photon transfer function that strongly deviates from the exact numerical result in the photon dominated regime, but closely tracks the exact solution in the ALP-dominated regime. The reason that this deviation in the photon dominated regime is not seen in reference~\cite{Mirizzi:2009} is that they derive a differential equation linear in $\Gamma l$. This reproduces the expected exponential absorption by integrating equation~\eqref{StatAveragedEqsTg}. While this exponential decay is correctly reproduced by equation~\eqref{SolForTransferFunctions} using the exact coefficients, invoking the approximation to obtain a differential equation yields the wrong absorption coefficient in equation~\eqref{StatAveragedEqsTg}. Fortunately, the area under the photon transfer function  in the photon dominated regime that is obtained with equations~\eqref{SolForTg} and ~\eqref{StatAveragedEqsTg} is such that the same value for the ALP population is produced. Asymptotically, both approaches therefore yield similar photon transfer functions.

In order to estimate the asymptotic value as well as the slope of the photon transfer function in the ALP-dominated regime we next solve
equation~\eqref{StatAveragedEqs} analytically, assuming that the coefficients are the same across all domains.
While this situation is not necessarily realized in the realistic astrophysical environment, this approximation is good
enough to obtain a rather accurate estimate. Introducing  $y\equiv \langle \mathcal{P}_{a\gamma}(l)\rangle z/l$
and $\alpha\equiv(1-\mm{exp}[-\Gamma l]+\langle \delta_\gamma(l)\rangle)/(2\langle \mathcal{P}_{a\gamma}(l)\rangle)\approx
\Gamma l/\langle \mathcal{P}_{a\gamma}(l)\rangle$ \cite{Mirizzi:2009}, as well as (typically very
small) $\beta=\langle \delta_a(l) \rangle/\langle \mathcal{P}_{a\gamma}(l)\rangle$ we can write the solution in a compact form
\begin{align}
  \label{TgammaStatAv}
  T_\gamma & = \frac{1+\beta-\nu+\kappa}{2\kappa}e^{-(\nu-\kappa)y}-
  \frac{1+\beta-\nu-\kappa}{2\kappa}e^{-(\nu+\kappa)y}\,,
\end{align}
where $\nu=3/4+(\alpha+\beta)/2$  and $\kappa=(\nu^2 -\alpha-\alpha\beta-\beta/2)^\frac12$. For
$\langle \mathcal{P}_{a\gamma}\rangle\ll \Gamma l \ll 1 $ (limit of strong absorption in the language of
reference~\cite{Mirizzi:2009}) the asymptotic solution is well approximated by
\begin{align}
  \label{TgammaApprox}
  T_\gamma \approx e^{-\Gamma z}+
  \frac{\langle \mathcal{P}_{a\gamma}\rangle^2}{2(\Gamma l)^2}e^{-\frac{\langle \mathcal{P}_{a\gamma}\rangle}{l}z}\,.
\end{align}
The first term in equation~\eqref{TgammaApprox} corresponds to photon absorption on the EBL. Thus,
the slope of $T_\gamma$ in the photon-dominated regime is given by $\Gamma$. On the other hand, the second term
comes from the ALPs that repopulate the photons in the ALP-dominated regime where most of the initial photons have been absorbed. Thus, the slope of $T_\gamma$ in the
ALP-dominated regime is given by $\langle \mathcal{P}_{a\gamma}\rangle/l$, and we conclude that the existence of ALPs leads to a much weaker asymptotic photon absorption.

\paragraph{Evolution equation for non-constant domain lengths.} Up to now we have assumed that the domain length is the same for all the domains along the line of sight. However, this assumption is probably not realized in nature and the domain length fluctuates following some kind of probability distribution. If one knows this distribution, one may attempt to include this information into equations~\eqref{SolForTransferFunctions} and~\eqref{StatAveragedEqs} by averaging the coefficients not only over $\vec{B}$ but also over $l$, and by replacing the left-hand side of equation~\eqref{StatAveragedEqs} according to $l \partial_z T_\gamma(z) \rightarrow \langle l \rangle \partial_z T_\gamma(z)$.

Such a generalization yields a good estimate of the asymptotic value for the photon transfer function obtained with Monte Carlo simulations. However, this agreement is very non-trivial: equations~\eqref{SolForTransferFunctions} and~\eqref{StatAveragedEqs} actually describe the expectation value for the transfer functions after a certain number of domains, while Monte Carlo simulations yield the expectation value at a particular distance from the source. When $l=\mm{const.}$ these two statements are equivalent. When $l$ is allowed to fluctuate, however, this agreement in not automatic but results from the very weak dependence of the photon transfer function on the distance in the ALP dominated regime. In this regime, distance from the source and transversing a corresponding mean number of domains are similar statements. In the photon dominated regime, where the transfer functions vary exponentially, the difference between solutions of equations~\eqref{SolForTransferFunctions} and~\eqref{StatAveragedEqs} and results of the Monte Carlo simulations is large.

\paragraph{Energy dependence of the transfer function.}

\begin{figure}[t!]
  \center
  \includegraphics[width=0.7\textwidth]{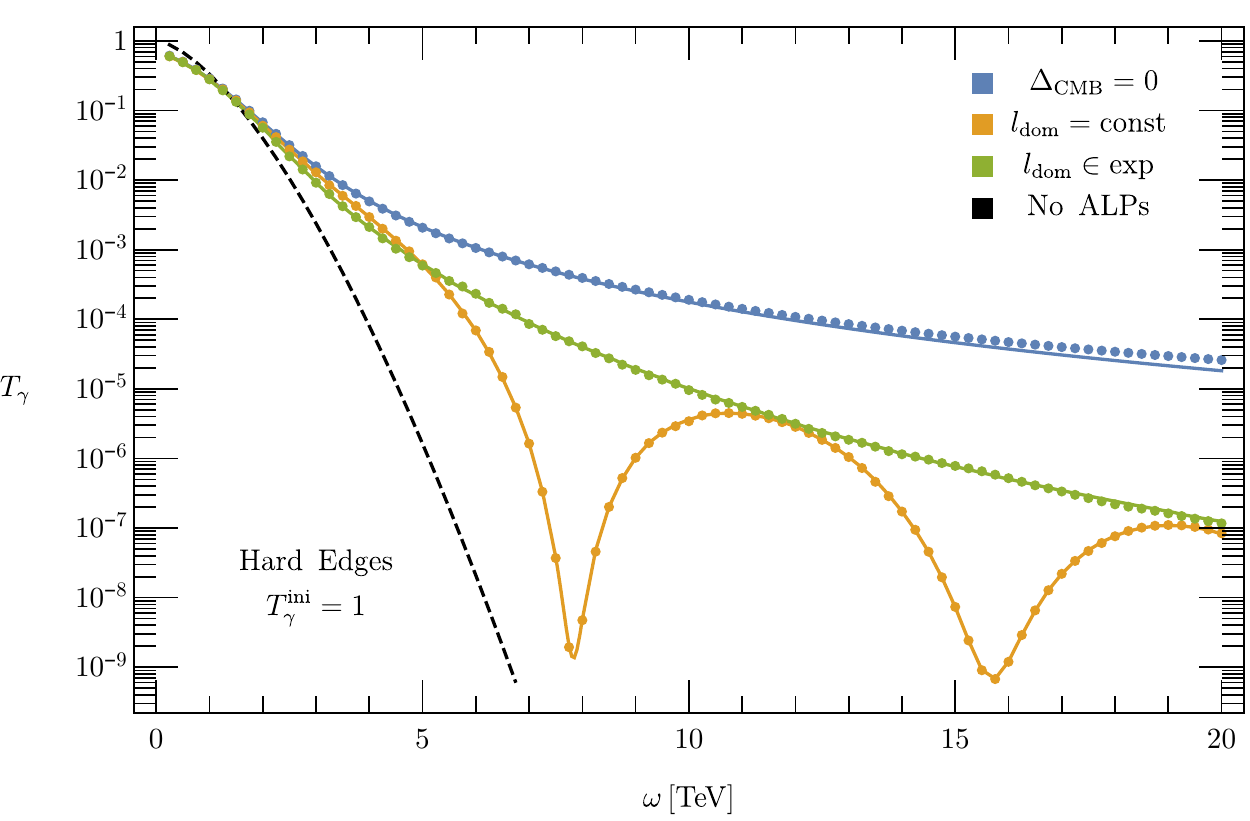}
  \caption{\label{Edependence}Energy dependence of the photon transfer function for magnetic field configurations with hard edges. In the absence of ALPs the photon population
    quickly decays, as is indicated by the black dashed curve. Results obtained in the presence of ALPs but neglecting photon-photon
    dispersion, i.e.\ in the case considered in reference~\cite{Mirizzi:2009}, for a constant domain length $l=10$~Mpc are shown in blue.
    They are to be contrasted with the results obtained using the same constant domain length but taking into account photon-photon
    dispersion shown in orange. The green curve draws the domain lengths from an exponential distribution with
    $\langle l\rangle=5\, \mm{Mpc}$ and also takes into account photon-photon dispersion. The dots represent results of the
    direct Monte Carlo simulations. This example has been produced assuming that the distance to the source is 1~Gpc and
  that the components of the magnetic field $B_x$ and $B_y$ are drawn from a Gaussian distribution with zero mean and standard deviation $\sqrt{2/3}$~nG.}
\end{figure}

The asymptotic solution \eqref{TgammaApprox} helps to understand the energy dependence of the photon transfer
function at large distances from the source presented in figure~\ref{Edependence}.
The energy dependence of the first term of equation~\eqref{TgammaApprox}, that describes the photon transfer function in
the absence of ALPs, is shown as a black dashed line. Because the absorption rate depends on the photon energy, this standard contribution
depends on the energy as well. At TeV energies it behaves as $\mm{exp}(-\Gamma d)$ with $\Gamma$ given in equation~\eqref{GammaApprox}.

The blue solid line in figure~\ref{Edependence} indicates the solution of equation~\eqref{StatAveragedEqs} with the setup of reference~\cite{Mirizzi:2009}, i.e.\ we include ALPs, but neglect photon-photon dispersion and keep
only terms of the first order in the absorption. The qualitative behavior of the solution is well described by the second term of equation~\eqref{TgammaApprox}. Whereas the oscillation probability is practically energy-independent in the setup of reference~\cite{Mirizzi:2009}, the growing absorption rate in the denominator  makes sure that
the transfer function decreases with increasing energy. This conclusion is also supported by the result of a direct
Monte Carlo simulation, that is shown by the blue dots. The result of the simulation and the solution of
equation~\eqref{StatAveragedEqs} match well at low energies but we see a deviation at higher energies, where higher order corrections in the absorption become important.

Results obtained \emph{with} photon-photon dispersion are shown in orange, where the dots again represent the
result of a direct Monte Carlo simulation and the solid line represents the solution of equation~\eqref{StatAveragedEqs}.
At small energies, the blue and orange curves agree well with each other. The reason is that at low energies photon-photon dispersion as well as corrections of higher order in the absorption can be neglected and equation~\eqref{StatAveragedEqs} reproduces the results of  reference~\cite{Mirizzi:2009}. At $\omega \gtrsim 1$~TeV the deviations become more pronounced. At even higher energies we observe suppression of the photon transfer function by several orders of magnitude. The reason is that the transition probability per domain, equation~\eqref{OscProb}, is suppressed when the photon-photon dispersion leads to a large phase velocity difference of the photons and ALPs. The peak-like structures
at $\omega \approx 7.8 \, \mm{TeV}$ and $\omega \approx 15.6 \, \mm{TeV}$ correspond to the minima of the conversion probability that occur when $\Re\,\Delta_{\mm{osc}}l/2=n\pi$ with integer $n$. They are an artifact of the assumption that all
domains have the same length. When the oscillation length becomes equal to this domain size, the photon-ALP system restores its initial condition in each domain. No oscillations occur over large distances because this resonance is sustained in every domain due to the magnetic field's grid-like structure. The resonant behavior is slightly broken by the imaginary part of $\Delta_{\mm{osc}}$ that is induced by absorption.

This oscillating behavior is lifted once we take into account that the domain length differs from domain to domain.
An example is shown in figure~\ref{Edependence} in green, where the domain length is drawn from an exponential distribution with $\langle l \rangle =5$~Mpc. This value for $\langle l\rangle$ is chosen because equation~\eqref{StatAveragedEqs} is controlled by
the ratio $\langle\mathcal{P}_{a\gamma}\rangle/\langle l\rangle$. For $\langle l \rangle= 5$~Mpc this ratio has
the same value as for the configuration with $l=10 \, \mm{Mpc}=\mm{const.}$ in the low energy limit. Hence, by construction, the orange and green curves coincide well for $\omega \rightarrow 0$. At higher energies we see deviations. In particular, the peak-like structure disappears because the resonance condition is no longer sustained in every domain, i.e.\ $\langle\mathcal{P}_{a\gamma}\rangle$ does not assume minima at $\omega \approx 7.8 \, \mm{TeV}$ and $\omega \approx 15.6 \, \mm{TeV}$.

The take-away message from all these numerical results is that due to photon-photon refraction the photon transfer function at TeV-range energies strongly depends  on the model of the magnetic field. One needs a phenomenologically viable model for the coherence length of the magnetic field and cannot rely on a simple model with a constant domain size. In the next section we will demonstrate that once the oscillation length becomes comparable to the typical width of the transition region between the domains the photon-ALP propagation
becomes close to adiabatic, and the assumptions and approximations that enter the derivation of equations~\eqref{SolForTransferFunctions} and~\eqref{StatAveragedEqs} break down.

\paragraph{Variance.}

\begin{figure}[t!]
  \center
  \includegraphics[width=0.7\textwidth]{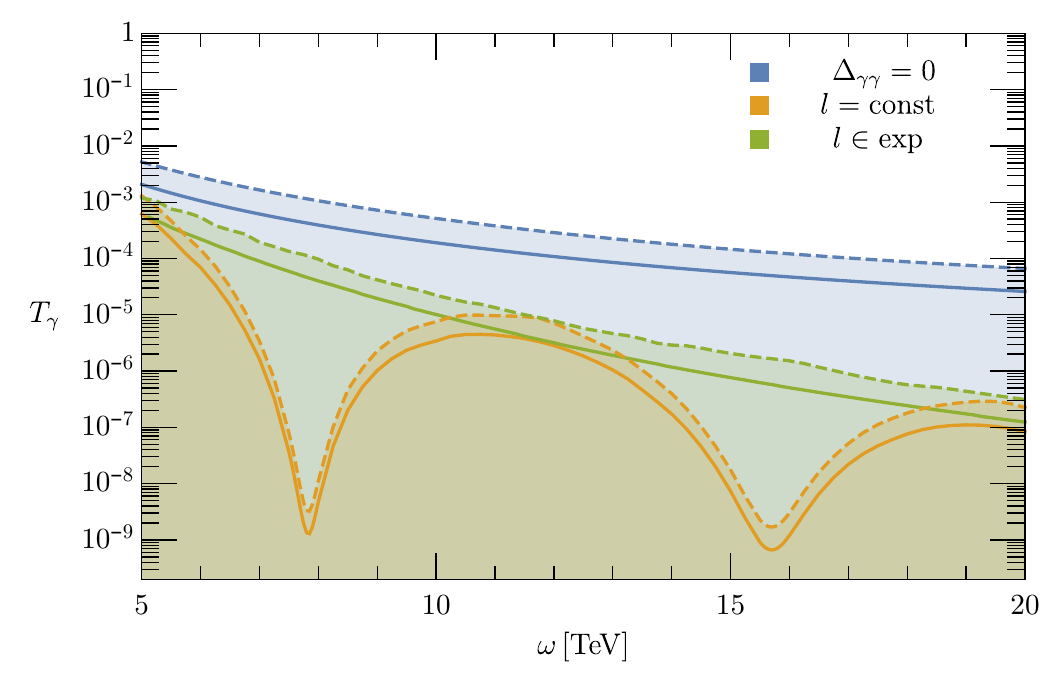}
  \caption{\label{fig:Variance}Energy dependence of the photon transfer function (solid lines) plus the corresponding
    standard deviation (dashed lines) computed using direct Monte Carlo simulations for energies above 5~TeV. The standard deviation is somewhat larger than the mean, so that, within one
    standard deviation around the mean, all values of $T_\gamma$ between zero and
    the dashed line can occur as indicated by the shading.
  The color-coding and the solid lines are the same as in figure~\ref{Edependence}.}
\end{figure}

To obtain a feeling for the spread of the photon transfer function around the mean  it is useful to compute the variance, $\delta T^2_\gamma=R_\gamma-T^2_\gamma $, where $R_\gamma \equiv \langle (\rho_{11}+\rho_{22})^2 \rangle$, and the standard deviation is ${\smash{(\delta T^2_\gamma)^{1/2}}}$.
However, one has to be careful in interpreting the variance because the probability distribution is not Gaussian.
The typical energy-dependence of the standard deviation computed using direct Monte Carlo
simulations is presented in figure~\ref{fig:Variance}, where we show the mean values (solid lines) and the mean plus standard deviation (dashed lines).
Note that the standard deviations are slightly larger than the corresponding mean values. Therefore, within one standard deviation around the mean, all values for the transfer function $T_\gamma$ between zero and the dashed line can occur as indicated by the shading. Still, figure~\ref{fig:Variance} suggests that different magnetic field scenarios lead to rather distinct
predictions even if one takes into account statistical fluctuations, and that, within one standard deviation, there is a clear
difference between the results
computed with and without taking into account photon-photon dispersion.

As has been shown in reference~\cite{Mirizzi:2009}, using equation~\eqref{EquationForRho} one can derive a coupled system
of differential equations that allows one to estimate $R_\gamma$, and consequently ${\smash{(\delta T^2_\gamma)^{1/2}}}$, without performing the Monte Carlo simulations. Within a single domain
\begin{align}
  \label{EqForRgamma}
  R_\gamma(z)-R_\gamma(0)&=\blangle \tfrac14\bigl(R^{11}_{11}(z)+R^{22}_{22}(z)\bigr)^2-1\brangle R_\gamma(0)
  +\blangle \mathcal{P}_{a\gamma}^2(z) \brangle R_a(0)\nonumber\\
  &+\tfrac18\blangle \bigl(R^{11}_{11}(z)-R^{22}_{22}(z)\bigr)^2\brangle (R_p(0)+\zeta_\gamma(0))+\blangle \mathcal{P}_{a\gamma}(z)  \bigl(R^{11}_{11}(z)+R^{22}_{22}(z)\bigr) \brangle \eta_{a\gamma}(0)\nonumber\\
  &
  -\blangle \bigl(I^{23}_{22}(z)\bigr)^2 \brangle \zeta_{a\gamma}(0)
  +\blangle \bigl(R^{23}_{22}(z)\bigr)^2 \brangle \zeta_{a\gamma 1}(0)
  +\blangle R^{23}_{22}(z)I^{23}_{22}(z) \brangle \zeta_{a\gamma 2}(0)\,,
\end{align}
where the expectation values $R_a\equiv \langle \rho^{\,2}_{33} \rangle$, $\eta_{a\gamma}\equiv
\langle ({\rho}_{11}+{\rho}_{22})\rho_{33} \rangle$, as well as $R_p\equiv \langle (\rho_{11}-{\rho}_{22})^2 \rangle$,
are combinations of the diagonals of the density matrix,
while the expectation values $\zeta_\gamma\equiv\langle ({\rho}_{12}+{\rho}_{21})^2 \rangle$ and
$\zeta_{a\gamma}\equiv\tfrac12\langle ({\rho}_{13}-{\rho}_{31})^2 \rangle+
\tfrac12\langle ({\rho}_{23}-{\rho}_{32})^2 \rangle$ are combinations of its off-diagonal elements.
To shorten the notation in equation~\eqref{EqForRgamma} we have introduced $R^{kl}_{ij}\equiv \Re(U_{ij}U^*_{kl})$ and
$I^{kl}_{ij}\equiv \Im(U_{ij}U^*_{kl})$, where $U_{ij}$ are elements of the evolution operator in equation~\eqref{EvolutionOperator}.
An expansion of the
last two coefficients in equation~\eqref{EqForRgamma}, $(R^{23}_{22})^2$ and $R^{23}_{22}I^{23}_{22}$, begins
with terms at least cubic in the components of the dispersive Hamiltonian matrix, see equation~\eqref{Hdis}.
The last two terms of
equation~\eqref{EqForRgamma} did not appear in the set of the six coupled differential equations derived in
reference~\cite{Mirizzi:2009} because only terms at most quadratic in the Hamiltonian were retained.
Once one resums terms of all orders, as has been done in the present work,
the set derived in reference~\cite{Mirizzi:2009} has to be extended by three additional equations
for  $\zeta_{\gamma 1}=\langle ({\rho}_{12}-{\rho}_{21})^2 \rangle$,
$\zeta_{a\gamma 1}=\tfrac12\langle ({\rho}_{13}+{\rho}_{31})^2 \rangle+ \tfrac12\langle ({\rho}_{23}+{\rho}_{32})^2 \rangle$,
and $\zeta_{a\gamma 2}=\tfrac{i}{2}\langle ({\rho}_{13}+{\rho}_{31})({\rho}_{13}-{\rho}_{31})\rangle+
\tfrac{i}{2}\langle ({\rho}_{13}-{\rho}_{31})({\rho}_{13}+{\rho}_{31})\rangle
+\tfrac{i}{2}\langle ({\rho}_{23}+{\rho}_{32})({\rho}_{23}-{\rho}_{32})\rangle+
\tfrac{i}{2}\langle ({\rho}_{23}-{\rho}_{32})({\rho}_{23}+{\rho}_{32})\rangle$.
The eight remaining (rather lengthy) differential equations that supplement equation~\eqref{EqForRgamma} are presented
in appendix~\ref{sec:variance}. These equations yield good estimates for the asymptotic variance that is computed using Monte Carlo simulations for the cases studied in figure~\ref{fig:Variance}.

\section{\label{sec:quasiadiabatic}Transition to quasi-adiabatic propagation at high energies}

In order to arrive at equation~\eqref{StatAveragedEqs} we had to assume that the magnetic field has a domain-like structure with abrupt transitions between the domains, i.e.\ that it has hard edges. If the extragalactic magnetic fields are created by, e.g., quasar outflows~\cite{Furlanetto:2001gx}, it seems reasonable that the inner part of a domain contains a magnetic field with approximately constant magnitude. Further outside, the magnetic field decreases or interacts with outflows from other quasars and we expect it to form a continuous profile. Realistic magnetic fields therefore have soft edges, i.e.\ they contain an additional length scale, the transition width $\lw$, that describes the distance over which the magnetic field continuously changes from its value in one domain to its value in the adjacent one. The oscillation length is bounded from above by $l_{\rm osc}\approx 200 \frac{10^{-11}\, {\rm GeV}^{-1}}{g_{a\gamma}}\frac{ {\rm nG}}{B}\, {\rm Mpc}$ and becomes smaller at high energies. The oscillation length will therefore be smaller than the transition width for large scale fields created before inflation, or will drop below the transition widths of typical magnetic fields at high energies if these fields are created by, e.g., quasar outflows. We therefore expect the solution of equation~\eqref{StatAveragedEqs} to start deviating from the result of a direct Monte Carlo simulation because the photon-ALP system begins to probe the structure of the magnetic field. In this section we provide a numerical example that confirms this theoretical expectation and derive analytically an approximate solution that describes quasi-adiabatic propagation of the photon-ALP system at high energies.

\paragraph{Comparable magnetic field configurations.}
\begin{figure}[t!]
  \center
  \includegraphics[width=0.7\textwidth]{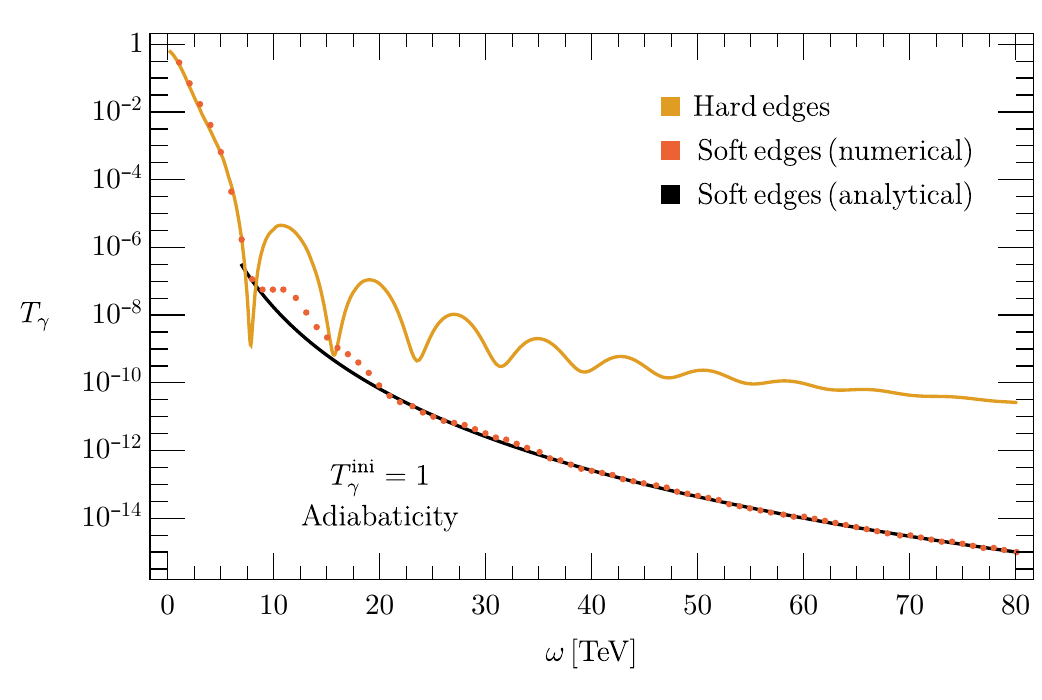}
  \caption{\label{fig:QAPropagation}Photon transfer functions computed using direct Monte Carlo simulation
    with magnetic field configurations featuring continuous transitions between the domains (red dots).
    At low energies it is well approximated by equation~\eqref{SolForTransferFunctions} (solid orange line), whereas
    at high energies it approaches the adiabatic solution equation~\eqref{TgApproxQAAveraged} (solid black line). This example has been
    produced assuming that distance to the source is 1~Gpc, that the domain size is $l=10$~Mpc, and that components
  of the magnetic fields $B_x$, $B_y$ are drawn from a Gaussian distribution with mean zero and standard deviation $\sqrt{2/3}$~nG.}
\end{figure}

For a given set of magnetic field configurations with soft edges we may obtain the mean photon transfer function by solving
equation~\eqref{EquationForRho} numerically. An example is shown in figure~\ref{fig:QAPropagation} with red dots.
A comparison of
this numerical result with the solution of equation~\eqref{StatAveragedEqs} or, equivalently, with the result of a Monte Carlo
simulation using magnetic field configurations with hard edges (solid orange line), is meaningful only if the two sets of magnetic fields are in some sense comparable.

\begin{figure}[t!]
  \center
  \begin{tabular}{cc}
    \includegraphics[width=0.45\textwidth]{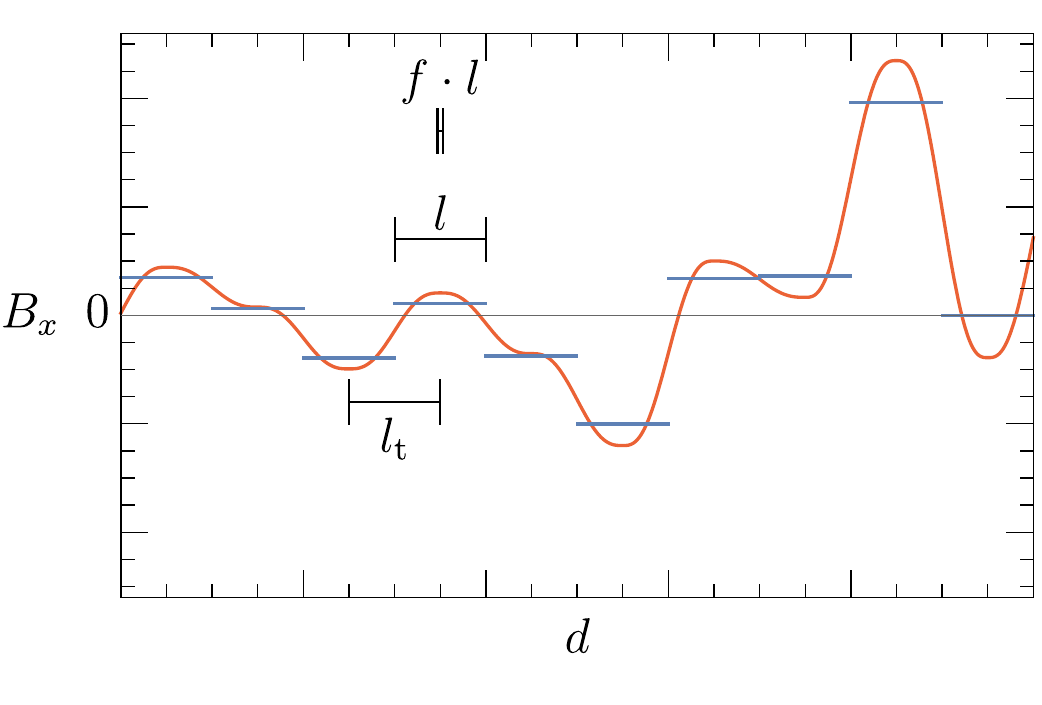}&\includegraphics[width=0.45\textwidth]{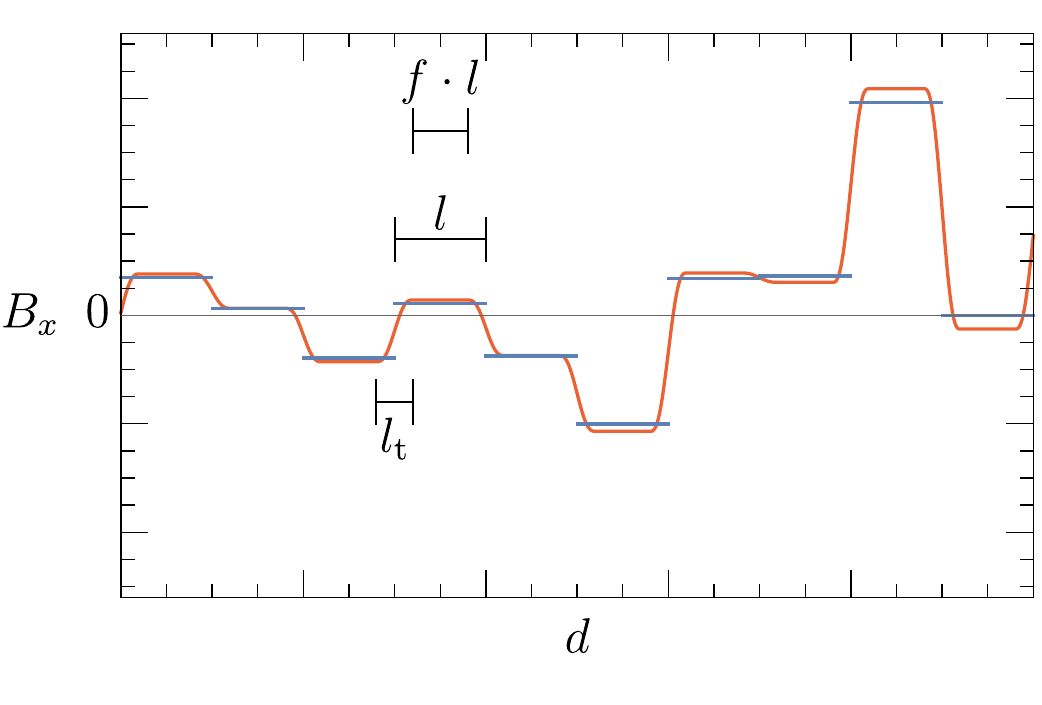}\\
    (a) & (b)
  \end{tabular}
  \caption{\label{fig:MagFields}Examples for magnetic fields with soft edges (red) derived from a magnetic field configuration with hard edges (blue). The filling factor $f$ on the left is zero, and on the right is it $0.6$. The domain length $l$ and the transition length $\lw$ are indicated.}
\end{figure}

From physical intuition we expect that the photon-ALP system does not probe the underlying structure of the magnetic field
as long as $l_{\rm osc}$ is much larger than $\lw$. This condition is fulfilled for vanishingly small ALP mass and $\omega \to 0 $ because whereas $l_{\rm osc}\rightarrow \infty$ in this limit, $\lw$ is bounded from above by the domain length, which is of the order of several $\mm{Mpc}$. In this limit the transition probability in each domain simplifies to
$P_{a\gamma}(l) = \bigl|\textstyle{\int}_{0}^{l}\D z \, \Delta_{a\gamma}(z)\bigr|^2$~\cite{Raffelt:1987im}.
We call two magnetic field configurations comparable if the oscillation probability in each domain is the same for the two
configurations. In particular, the example presented in figure~\ref{fig:QAPropagation} relies on the following procedure to construct normalized magnetic fields. We generate a magnetic field configuration with hard edges for the two transversal directions, which we then interpolate separately with a continuously varying magnetic field. In the center of each domain we place a constant subdomain with a size that is given by the filling factor $f\equiv 1-\lw/l$ times the original domain size, see figure~\ref{fig:MagFields}. We require the boundaries between each subdomain and the outer magnetic field structure to be reasonably smooth, which we enforce by setting the first and second derivatives to zero at the boundary. This provides us with six conditions for each interpolation between two subdomains. We choose a fifth order polynomial between each subdomain.

This procedure is in contrast to the one developed in reference~\cite{Wang:2015dil} where the authors take a magnetic field with constant magnitude  but varying azimuth angle, which can be a function of distance. They match the magnetic field direction at the border of each domain. In general this matching leads to conversion probabilities in each domain that are different for scenarios with hard and soft edges, as is also observed by the authors of reference~\cite{Wang:2015dil}. This finding is however not surprising, because the two magnetic field realizations are not comparable.

Getting back to figure~\ref{fig:QAPropagation}, at low energies the photon transfer function obtained using a direct Monte Carlo
simulation with hard edges and soft edges agree on the sub-percent level, by construction.  Pronounced deviations arise once $l_\mm{osc} \approx \lw$, which happens at $\omega \approx 8 \, \mm{TeV}$. The ALP-photon system becomes sensitive to the structure of the magnetic field. We also compared the photon transfer functions for different transition widths and
found that the photon transfer functions vary continuously between the cases $\lw=0$ and $\lw =l$.

At even higher energies and finite $\lw$, the ALP-photon system starts adopting quasi-adiabatically to the changes of the background because the oscillation length is much smaller than any variation of the magnetic field. This case is studied in
the remainder of this section. Our result extends the work of reference~\cite{Wang:2015dil} in that we include absorption which qualitatively changes photon-ALP propagation in the quasi-adiabatic regime.

\paragraph{Slowly varying magnetic field.} If the magnetic field varies slowly everywhere, the evolution of the photon-ALP system becomes quasi-adiabatic and is conveniently described in the basis of propagation eigenstates, $\mathbf{c}=\mathbf{O} \mathbf{A}$, with
\begin{align}
  \mathbf{O}=
  \begin{pmatrix}
    1 & 0 & 0\\
    0 & c_\alpha & s_\alpha \\
    0 & -s_\alpha & c_\alpha
  \end{pmatrix}
  \begin{pmatrix}
    c_\phi & s_\phi & 0\\
    -s_\phi & c_\phi & 0 \\
    0 & 0 & 1
  \end{pmatrix}\,,
\end{align}
where $\phi$ is the angle between the direction of the (linear) photon polarization and the direction of the magnetic field, and $\alpha$ is
the photon-ALP mixing angle introduced in equation~\eqref{MixingAngle}. Taking the derivative of $\mathbf{c}$ and using equation~\eqref{SchroedEq}
together with the orthogonality relation $\mathbf{O}^T \mathbf{O} = \mathbf{O} \mathbf{O}^T = \mathbf{1}$ we arrive at
\begin{align}
  \label{PropBasisGeneral}
  i\partial_z
  \begin{pmatrix}
    c_1 \\
    c_2 \\
    c_{a} \\
  \end{pmatrix}=
  \begin{pmatrix}
    E_1 & ic_\alpha\partial_z \phi & -is_\alpha \partial_z\phi \\
    -ic_\alpha\partial_z \phi & E_2 & i\partial_z\alpha \\
    is_\alpha \partial_z\phi & -i\partial_z\alpha & E_3
  \end{pmatrix}
  \begin{pmatrix}
    c_1 \\
    c_2 \\
    c_a \\
  \end{pmatrix}\,,
\end{align}
where $E_i$ are the adiabatic eigenvalues of  $\mathbf{H}$ calculated in section~\ref{sec:averaged}. The
derivatives of $\phi$ and $\alpha$ in the off-diagonals of equation~\eqref{PropBasisGeneral} originate from $\partial_z\mathbf{O}$.
The adiabatic limit corresponds to $\partial_z\phi \rightarrow 0$ and $\partial_z\alpha \rightarrow 0$, i.e.\ to the propagation in a constant magnetic field.
As follows from equation~\eqref{PropBasisGeneral}, in this limit each of the propagation eigenstates evolves
independently. In other words, if the propagation is adiabatic and the system begins its evolution
in one of its propagation eigenstates then it remains in the same eigenstate in the course of its evolution,
also in the presence of absorption.

The degree of deviation from adiabaticity depends on the size of the off-diagonals elements of the
Hamiltonian matrix in equation~\eqref{PropBasisGeneral} relative to the size of its diagonal elements.
The smoother the transition from one domain to another is, the smaller $\partial_z\phi$ and $\partial_z\alpha$ are, and the closer is the photon-ALP propagation to adiabatic.
For the field configurations considered in reference~\cite{Mirizzi:2009} the magnetic field abruptly changes at the domain borders and the
derivatives of $\phi$ and $\alpha$ are infinite. Therefore, in this setup the photon-ALP propagation
at the domain borders is always non-adiabatic.

For a given configuration of the magnetic field, i.e.\ for given $\partial_z\phi$ and $\partial_z\alpha$,
the degree of non-adia\-ba\-ticity depends on the size of the diagonal elements $E_i$. That is, low-energy photons tend to propagate non-adiabatically, while high energy photons tend to adapt to the magnetic field adiabatically. This implies that at sufficiently high energies using magnetic field configurations with hard edges becomes too crude an approximation and the evolution equation~\eqref{StatAveragedEqs} cease to be applicable.

\paragraph{Limit of decoupled ${\bm{c_1}}$.}

For slowly varying  $\alpha$, $\partial_z\alpha$, and $\partial_z\phi$, equation~\eqref{PropBasisGeneral} can be solved
approximately.
Before analyzing the general case it is instructive to first study a simplified setup in which only one photon polarization mixes with the ALP. This is the case if the azimuthal angle of the magnetic field is constant, $\phi=\rm const$. In this case the component  $c_1$
decouples, see equation~\eqref{PropBasisGeneral}. Given the existing constraints on the photon-ALP coupling, see e.g.\ reference~\cite{Dias:2014osa}, for TeV-range
photon energies and the typical assumptions about the magnetic field, $B\lesssim 10^{-9}\ \rm{G}$, the contribution of the magnetic field
to $E_{1,2}$ is subdominant. Therefore at these energies the real part of $E_{1,2}$ is dominated by the dispersion on the CMB and the imaginary part
by the EBL absorption, which are independent of the spatial coordinate $z$ if redshift is neglected. With these approximations the solution for $c_1$ reads $c_1(z)=\exp(-iE_1z)c_1(0)$.

If $\partial_z\alpha$ were constant, then the
system of coupled differential equations for $c_2$ and $c_a$ could be solved analytically,
\begin{align}
  \label{QAAnalytSol}
  \begin{pmatrix}
    c_2 (z)  \\
    c_a (z)
  \end{pmatrix}=
  \left[
    \begin{pmatrix}
      1 & i\eta \\
      -i\eta & \eta^2
    \end{pmatrix}\frac{\exp(-i\mathcal{E}_2 z)}{1+\eta^2}
    +
    \begin{pmatrix}
      \eta^2 & -i\eta \\
      i\eta & 1
    \end{pmatrix}\frac{\exp(-i\mathcal{E}_3 z)}{1+\eta^2}
  \right]
  \begin{pmatrix}
    c_2 (0)  \\
    c_a (0)
  \end{pmatrix}\,,
\end{align}
where $\eta\equiv \partial_z\alpha/(E_2-\mathcal{E}_3)$, and $\mathcal{E}_{2,3}$ are eigenvalues of the Hamiltonian matrix in
equation~\eqref{PropBasisGeneral} that in the limit $\partial_z\alpha\rightarrow 0$ revert to $E_2$ and $E_3$ respectively.
Although linear growth of the mixing angle over astrophysically large distances is not physical, equation~\eqref{QAAnalytSol} can
nevertheless be used to understand the qualitative behavior of the photon and ALP transfer functions. In the cases of physical
interest, the photon-ALP system is produced (approximately) in a photon interaction eigenstate. Thus, because the mixing angle is typically very small, we have
$c_2(0) \gg c_a(0)$. For these initial conditions and at small distances from the source
\begin{subequations}
  \label{SmallZExp}
  \begin{align}
    c_2(z) & \approx \exp(-iE_2 z)c_2(0)\,,\\
    c_a(z) & \approx i\eta \left[1-\exp(-iE_2z)\right]c_2(0)\,,
  \end{align}
\end{subequations}
where we have used $E_3\ll E_2$. In other words, at small distances from the source the photon population
exponentially decays due
to absorption on the EBL. At the same time it sources the ALP population that rapidly grows as $c_a(z)\approx - z \partial_z
\alpha\, c_2(0)$. On the other hand, at large distances from the source $\exp(-iE_2z)\ll \exp(-iE_3z)$
and the large-$z$ asymptotics is given by
\begin{subequations}
  \label{LargeZExp}
  \begin{align}
    c_2(z) & \approx \eta^2\exp(-iE_3z)c_2(0)\,,\\
    c_a(z) & \approx i\eta \exp(-iE_3z)c_2(0)\,.
  \end{align}
\end{subequations}
For ALPs the transition
from the small to the large-$z$ regime happens when $|[i\eta+c_a(0)]
\exp(-iE_3z)|$ becomes much larger than  $|-i\eta\exp(-iE_2z)|$, i.e.\ almost immediately. On the other hand, for photons this transition happens when $\exp(-iE_2z)$
becomes comparable to $|[\eta^2-i\eta c_a(0)]\exp(-iE_3z)|$, i.e.\ quite far away from the source.

Guided by equation~\eqref{QAAnalytSol} and its expansion at small \eqref{SmallZExp} and large~\eqref{LargeZExp}
distances from the source we may now derive an approximate analytic solution valid for slowly varying
$\partial_z\alpha$ assuming again that the system is produced in the photon propagation eigenstate, or a state
close to it. Because $c_2(0) \gg c_a(0)$ for small $z$ one may neglect the $i\partial_z\alpha$
term in the equation for $c_2$ (the (2,3) entry of the Hamiltonian matrix). In this approximation
the small-$z$ solution for $c_a(z)$ reads
\begin{align}
  \label{QASolSmallz}
  c_a (z)\approx \bigl[c_a(0)-c_\gamma(0)\textstyle{\int}_0^z \D z'\, \Delta_\mm{osc}\eta(z')\exp\bigr(-i\Delta_\mm{osc} z'\bigr) \bigr]e^{-i\textstyle{\int}_0^z E_3 \D z''}\,,
\end{align}
with $\Delta_\mm{osc} = E_2-E_3$ taken constant in $z$,   whereas $\eta\approx \partial_z\alpha/\Delta_\mm{osc}$ now depends on $z$. Because $E_3\approx -\alpha^2
E_2^*$ (as can be verified by expanding equation~\eqref{DeltaDef} in powers of $\Delta_{a\gamma}$), $E_3$ can in principle strongly
depend on $z$ and therefore $\exp(-i\textstyle{\int}_0^z E_3 \D z')$ is not necessarily well approximated by $\exp(-iE_3z)$.

The integral in equation~\eqref{QASolSmallz} can be estimated in the quasi-adiabatic limit by using the Fourier representation
\begin{equation}
  \Delta^2_{\rm osc}\eta \approx \partial_z \Delta_{a\gamma}(z) = \int_{-\infty}^{\infty}\frac{{\D} k}{2\pi}\, ikf(k)e^{ikz}\,.
\end{equation}
Performing the spatial integral, we obtain for the  ALPs in the small-$z$ limit
\begin{equation}
  c_a(z)\approx \left[c_a(0)-c_\gamma(0)\int_{-\infty}^{\infty}\frac{{\D} k}{2\pi}\,\frac{ kf(k) }{\Delta_{\mm{osc}} (\Delta_\mm{osc}-k)}\right]e^{-i\textstyle{\int}_0^z E_3 \D z'}\,.
\end{equation}
If $f(k)$ falls off fast enough, i.e. if the power spectrum  of the magnetic field is dominated by large scales, the dominant contribution to the $k$-integral can be estimated to be $i\eta(0)c_\gamma(0)$. This approximation is meaningful
as long as the contribution from small scales, that stems from the integration ranges $(-\infty,-\Re\Delta_{\rm osc})\cup
(\Re\Delta_{\rm osc},\infty)$, is small compared to the contribution from large scales.

Because for ALPs the transition from the small-$z$ to the large-$z$ regime happens rather quickly,  one can use the
solution for the ALP-like propagation eigenstate $c_a$ also at large distances from the source. On the other hand,
at large $z$ the photon amplitude decays and the (2,3) entry of the Hamiltonian matrix becomes important. Solving the equation for
$c_2(z)$ with $c_a(z)$ (as given by equation~\eqref{QASolSmallz}) as a source we obtain at large distances from
the source
\begin{align}
  \label{QAApproxSol}
  \begin{pmatrix}
    c_2 (z)  \\
    c_a (z)
  \end{pmatrix} & \approx
  \exp\bigl(-i\textstyle{\int}_0^z\D z'\, E_3 \bigr)
  \begin{pmatrix}
    \eta(0)\eta(z) & -i\eta(z) \\
    i\eta(0) & 1
  \end{pmatrix}
  \begin{pmatrix}
    c_2 (0)  \\
    c_a (0)
  \end{pmatrix}\,,
\end{align}
which generalizes the second term of equation~\eqref{QAAnalytSol} to the case of a slowly varying $\partial_z\alpha$.

The parametrical dependence of $c_2(z)$ on $\eta$  can be qualitatively understood as follows. The contribution
of $c_2$ into the buildup of the ALP-like eigenstate $c_a$ is proportional to $\partial_z\alpha$. At large
distances from the source the back-conversion amplitude of $c_a$ into $c_2$ is proportional to $\partial_z
\alpha$ as well. This explains why the (1,1) element of the matrix in equation~\eqref{QAApproxSol} is parametrically of the order
of $\eta^2$. In addition, there is a contribution proportional to the initial abundance of the ALP-like propagation eigenstate,
$c_a(0)$, which is of the order of  $\partial_z\alpha\, c_a(0)$. This explains why the (1,2) element of the
matrix in equation~\eqref{QAApproxSol} is parametrically of the order $\eta$.

The $c_1$ component, which is decoupled from the ALPs, decays as $\exp(-i E_1 z)$ and therefore is essentially
zero at large distances from the source. Taking this into account and rotating back to the interaction basis we find
for the asymptotic value of the photon transfer function
\begin{align}
  \label{TgApproxQA}
  T_\gamma\approx \left|A_2(0)\right|^2\left|s_{\alpha}(0)-i\eta(0)c_\alpha(0)\right|^2\left|s_{\alpha}(z)+i\eta(z)c_\alpha(z)
  \right|^2 \exp\bigl(2\textstyle{\int}_0^z\, \Im E_3\, \D z' \bigr)\,.
\end{align}
The first term, $\left|A_2(0)\right|^2$, encodes the initial conditions and for pure photon initial condition can be
parameterized as $c_\phi^2(0)$, see e.g.\ equation~\eqref{InitialState}. Similarly to the case of adiabatic propagation
without absorption, the result depends on the conditions at the source (the second terms) and the detector (the third
term). The absorption manifest itself directly in the slowly decaying ($\Im E_3 < 0$) exponential factor that ``measures'' the
degree to which the ALPs are exposed to absorption due to mixing with photons.

\paragraph{General case.}

In the general case, the component $c_1$ also does not contribute to the photon transfer function if we keep in the asymptotic amplitude $\bm{c}(z)$ terms at most quadratic in the small quantities $\partial_z\alpha$, $\partial_z\phi$, and $s_\alpha$. This is seen already from the analytic solution of equation~\eqref{PropBasisGeneral}
that can be obtained for (approximately) constant  $\alpha$, $\partial_z\alpha$ and $\partial_z\phi$,
\begin{align}
  \label{QAApproxSolGeneral}
  c_1(z)\sim A_2(0)\xi[s_\alpha-i\eta c_\alpha][s_\alpha+i\eta c_\alpha]
  \exp\bigl(-i \textstyle{\int}^z_0 \mathcal{E}_3\, \D z'\bigr)+{\rm higher~order~terms}\,,
\end{align}
where $\xi \approx \partial_z\phi/E_2$. Evidently, at large distances from the source the expansion of $c_1(z)$
begins with terms cubic in the small quantities.

The qualitative arguments presented in the previous paragraph can be used to demonstrate that equation~\eqref{QAApproxSolGeneral} is also valid for small and slowly varying $\alpha$, $\partial_z\alpha$ and $\partial_z\phi$.
Let us first consider the contribution of $c_2$ and $c_1$ into the
buildup of the ALP-like eigenstate $c_a$. Whereas  for $c_2$ the conversion
amplitude is proportional to $\partial_z\alpha$, for $c_1$ the direct conversion amplitude is proportional to
$s_\alpha\partial_z\phi$, i.e.\ is quadratic in the small quantities. In addition, $c_1$ can be converted into
$c_2$ with the amplitude proportional to $c_\alpha \partial_z\phi$, and subsequently into $c_a$ with the
amplitude proportional to $\partial_z\alpha$. Thus, this indirect conversion amplitude is proportional to $c_\alpha
\partial_z\phi \partial_z\alpha$, i.e.\ it is again of second order in the small quantities. Next we consider the back
conversion of $c_a$ into $c_2$ and $c_1$ at large distances from the source. From the presented
arguments it follows that, to leading order, the dynamically induced ALP interaction eigenstate population given by  $c_a$ is proportional to
$\partial_z\alpha$. A direct back-conversion into
$c_2$ yields another factor $\partial_z\alpha$ and we recover the $\eta^2$ dependence that we observed in the preceding
subsection, see equation~\eqref{QAApproxSol}. A direct back-conversion into $c_1$ on the other hand yields the factor $s_\alpha\partial_z\phi$ and
we end up with $c_1$ being proportional to $s_\alpha\partial_z\phi\,\partial_z\alpha$, i.e.\ cubic in the small
parameters. In addition, there are contributions proportional to the initial abundance of the ALP-like eigenstate,
$c_a(0)\sim s_\alpha$. For $c_2$ the resulting contribution is then of the order of $s_\alpha \partial_z\alpha$.
On the other hand, for $c_1$ the resulting contribution is of the order of $s_\alpha^2\partial_z\phi$, i.e.\ it is
again cubic in the small parameters. Similarly, the amplitude of the indirect conversion of $c_a(0)$ into
$c_2$ and subsequently into $c_1$ is proportional to $c_\alpha s_\alpha \partial_z\alpha\,\partial_z\phi$
and is again of third order. Therefore, in the quasi-adiabatic regime to second order in the small quantities
$\partial_z\alpha$, $\partial_z\phi$, and $s_\alpha$, the $c_1$ component is decoupled from the evolution of
$c_2$ and $c_a$ and the result reverts to equation~\eqref{TgApproxQA}.

Even for the most distant gamma-ray sources known to date the factor $\exp(\textstyle{\int}_0^z\, \Im E_3\, \D z')$ is well approximated
by unity. Assuming an unpolarized source, averaging of the factor $|A_2(0)|^2$ yields $\sfrac12$. Finally, because configurations of the magnetic field at the source and the detector are not correlated, we obtain for the statistically averaged photon transfer
function
\begin{align}
  \label{TgApproxQAAveraged}
  T_\gamma\approx \tfrac12\langle\left|s_{\alpha}(0)-i\eta(0)c_\alpha(0)\right|^2\rangle
  \langle\left|s_{\alpha}(z)+i\eta(z)c_\alpha(z)\right|^2\rangle \,.
\end{align}
For the parameters used in figure~\ref{fig:QAPropagation}, the approximate solution equation~\eqref{TgApproxQAAveraged} is expected to
become applicable at energies of the order of 8~TeV, where the oscillation length becomes smaller than the chosen value $\lw=l=10$ Mpc. As can
be inferred from figure~\ref{fig:QAPropagation}, between 8~TeV and 20~TeV the agreement between the numerical and
analytical results is qualitative at best. At energies larger than 20~TeV the approximate solution equation~\eqref{TgApproxQAAveraged} finally becomes accurate.

We would like to emphasize that the ``hard'' and ``soft'' edge field configurations in figure~\ref{fig:QAPropagation}
can hardly be considered as realistic and serve merely as benchmarks corresponding to the extreme cases of a non-adiabatic
and close-to-adiabatic propagation respectively. For more realistic magnetic field configurations we expect the transfer
function of TeV-energy photons to lie somewhere between the two extremes. For these intermediate energies precise results can only be obtained numerically.  On the other hand, at high energies the propagation necessarily becomes close to adiabatic and the
resulting transfer function approaches the ``soft edge'' numerical curve, which depends only on the magnetic field configuration at
the source and the detector.

\section{\label{sec:conclusions}Conclusions}

It has recently been realized that a photon gas is a dispersive medium for photon propagation.
This otherwise tiny effect dominates the dispersion of TeV gamma rays and, while it used to be ignored, can modify the oscillation between TeV gamma rays and axion-like particles in
astrophysical magnetic fields. In the present work we have studied the impact of this effect
on the photon-ALP propagation in extragalactic magnetic fields at TeV energies relevant for CTA. For conceptual clarity we have neglected redshift and have used a rather crude approximation for the photon absorption rate. We have identified two important effects that are inevitable for TeV gamma rays mixing with ALPs.

First, photon-photon refraction increases with energy and thereby causes a photon-ALP mixing angle that decreases with increasing
energy: the phase velocity difference between photon and ALP interaction eigenstates grows approximately linearly with energy, while the term mixing the two eigenstates remains unaffected.  A smaller mixing angle means that a smaller fraction of the initial photons is
converted into ALPs and later back-converted into photons close to the detector. In other
words, photon-photon refraction results in a photon flux that for large energies is asymptotically smaller than that expected in the usually studied
case of maximal mixing. In order to compute the photon transfer
function in the presence of photon-photon refraction we have generalized the differential equation formalism developed
by Mirizzi and Montanino to the case of arbitrary mixing angles by resumming terms of all orders in the
dispersive and absorptive Hamiltonian.

Second, photon-photon refraction results in a photon-ALP oscillation length decreasing with increasing
energy.  The reason is again that the phase velocity difference between photon and ALP interaction eigenstates, which determines the
oscillation frequency and the oscillation length, grows approximately linearly with energy. An important
implication of this effect is that at CTA energies the photon-ALP oscillation length becomes comparable
or smaller than the typical length scales associated with extragalactic magnetic fields, so that the photon-ALP
propagation becomes very sensitive to the exact structure of the magnetic field. In particular, the
simplified model of the extragalactic magnetic fields adopted in many previous publications becomes too crude
at these energies. At even higher energies the propagation is close to adiabatic and, as we have demonstrated, the photon transfer function
depends on the magnetic field and its first derivative at the source and the detector but is not sensitive to the intermediate magnetic field
configuration.

While current Cherenkov telescopes have limited sensitivity to TeV gamma rays from distant blazars, with the advent of CTA, HAWC, and
HiSCORE, future telescopes will be much more sensitive to these high energy photons.  This increased sensitivity will prove important to
decide if the alleged discrepancy between different measurements of the opacity of extragalactic space persists. Our results are crucial for the
interpretation of blazar spectra in the TeV range in terms of photon-ALP oscillations. While we have shown that photon-photon refraction
typically suppresses the amount of TeV gamma rays expected in comparison to the maximal mixing scenario and drives
it closer to that expected in the standard scenario, quantitative predictions for e.g.\ CTA require refined models of the extragalactic
magnetic field configurations. However, precisely because photon-photon refraction diminishes the impact of
photon-ALP mixing, the most important contribution to photon-ALP conversions that
could explain the larger than expected transparency of the Universe to TeV gamma rays probably occurs in the sources and the Milky Way, where typical magnetic fields overpower photon-photon refraction. In this case, the qualitative insights of the present work apply to scenarios with non-negligible ALP masses.

\newpage
\acknowledgments

We acknowledge partial support by the Deutsche Forschungsgemeinschaft under the
Excellence Cluster ``Universe'' (Grant No.\ EXC 153) and by the Horizon 2020
Marie Sk\l{}odowska-Curie Actions of the European Union under the Innovative
Training Network ``Elusives'' (Grant No.\ H2020-MSCA-ITN-2015/674896-ELUSIVES).
H.V. was partially supported by the Department of Energy, under contract
DE-AC02-76SF00515.

\appendix

\section{\label{sec:variance}Variance}
As has been argued in section~\ref{sec:averaged}, a rough estimate of the variance can be obtained by solving
a set of nine coupled differential equations. Within a single domain we obtain for  $R_\gamma= \langle (\rho_{11}+\rho_{22})^2 \rangle$,
$R_a=\langle \rho^{\,2}_{33} \rangle$, and $\eta_{a\gamma}=\langle ({\rho}_{11}+{\rho}_{22})\rho_{33} \rangle$
\begin{subequations}
  \label{EqsForRaandetaagamma}
  \begin{align}
    \Delta R_\gamma&=\blangle \tfrac14\bigl(R^{11}_{11}+R^{22}_{22}\bigr)^2-1\brangle R_\gamma
    +\blangle \mathcal{P}_{a\gamma}^2 \brangle R_a
    +\tfrac18\blangle \bigl(R^{11}_{11}-R^{22}_{22}\bigr)^2\brangle (R_p+\zeta_\gamma)\nonumber\\
    &+\blangle \mathcal{P}_{a\gamma}  \bigl(R^{11}_{11}+R^{22}_{22}\bigr) \brangle \eta_{a\gamma}
    -\blangle \bigl(I^{23}_{22}\bigr)^2 \brangle \zeta_{a\gamma}
    +\blangle \bigl(R^{23}_{22}\bigr)^2 \brangle \zeta_{a\gamma 1}
    +\blangle R^{23}_{22}I^{23}_{22} \brangle \zeta_{a\gamma 2}\,,\\
    \Delta R_a & = \tfrac14 \blangle \mathcal{P}^2_{a\gamma} \brangle R_\gamma
    +\blangle \bigl(R^{33}_{33}\bigr)^2-1 \brangle R_a
    +\tfrac18 \blangle \mathcal{P}^2_{a\gamma} \brangle (R_p+\zeta_\gamma)
    +\blangle \mathcal{P}_{a\gamma} R^{33}_{33} \brangle \eta_{a\gamma}\nonumber\\
    &- \blangle \bigl(I^{23}_{33}\bigr)^2\brangle \zeta_{a\gamma}
    + \blangle \bigl(R^{23}_{33}\bigr)^2\brangle \zeta_{a\gamma 1}
    - \blangle R^{23}_{33} I^{23}_{33}\brangle \zeta_{a\gamma 2}\,,\\
    \Delta \eta_{a\gamma} & =
    \tfrac14 \blangle \mathcal{P}_{a\gamma}\bigl(R^{11}_{11}+R^{22}_{22}\bigr) \brangle R_\gamma
    +\blangle \mathcal{P}_{a\gamma} R^{33}_{33} \brangle R_a
    -\tfrac18\blangle \mathcal{P}_{a\gamma}\bigl(R^{11}_{11}-R^{22}_{22}\bigr) \brangle \bigl(R_p+\zeta_\gamma\bigr)\nonumber\\
    & +\tfrac12\blangle R^{33}_{33}\bigl(R^{11}_{11}+R^{22}_{22}\bigr)+\mathcal{P}^2_{a\gamma}-2 \brangle \eta_{a\gamma}
    + \blangle I^{23}_{22} I^{23}_{33} \brangle \zeta_{a\gamma}
    + \blangle R^{23}_{22} R^{23}_{33} \brangle \zeta_{a\gamma 1}\nonumber\\
    &+\tfrac12 \blangle \mathcal{P}_{a\gamma} I^{33}_{22} \brangle \zeta_{a\gamma 2}\,.
  \end{align}
\end{subequations}
On the left-hand side of equations~\eqref{EqsForRaandetaagamma} we use  $\Delta R_\gamma \equiv R_\gamma(z) - R_\gamma(0)$,
etc., to shorten the lengthy expressions. For the same reason, on the right-hand side of equations~\eqref{EqsForRaandetaagamma} we
omit arguments of the expectation values $R_\gamma$, etc. , that are evaluated at the beginning of the domain, as well as arguments of
the coefficients $R^{kl}_{ij}=\Re(U_{ij}U^*_{kl})$ and $I^{kl}_{ij}=\Im(U_{ij}U^*_{kl})$,
that are evaluated at distance $z\leq l$ from the beginning of the domain, see e.g. equation~\eqref{EqForRgamma}.

As follows from equation~\eqref{EquationForRho}, the combination $R_\gamma+R_a+2\eta_{a\gamma}=\langle (\rho_{11}+\rho_{22}+\rho_{33})^2 \rangle=\langle (\tr \bm{\rho})^2 \rangle$ is conserved in the absence of photon absorption. In equation~\eqref{EqsForRaandetaagamma} this property is reflected in that the
right-hand side of this sum vanishes in each domain in the limit of vanishing absorption,
\begin{align}
  \Delta R_\gamma+\Delta R_a+2\Delta \eta_{a\gamma}= - 2\Gamma l (R_\gamma + \eta_{a\gamma})+\mathcal{O}(l^2)\,.
\end{align}
The evolution equations for another combination of the diagonals of the density matrix,
$R_p=\langle (\rho_{11}-{\rho}_{22})^2 \rangle$, reads
\begin{subequations}
  \begin{align}
    \Delta R_p & =
    \tfrac18 \blangle \bigl(R^{11}_{11}-R^{22}_{22}\bigr)^2\brangle R_\gamma
    +\tfrac12 \blangle \mathcal{P}^2_{a\gamma}\brangle R_a
    +\blangle \tfrac18 R^{22}_{11}\bigl(R^{11}_{11}+R^{22}_{22}\bigr)+\tfrac{3}{32}\bigl[ \bigl(R^{11}_{11}\bigr)^2
      + \bigl(R^{22}_{22}\bigr)^2\nonumber\\
    & +2 \bigl(R^{22}_{11}\bigr)^2 - 2\bigl(I_{11}^{22}\bigr)^2\bigr] + \tfrac38 R^{11}_{11} R^{22}_{22} -1 \brangle R_p
    - \tfrac12 \blangle \mathcal{P}_{a\gamma}\bigl(R^{11}_{11}-R^{22}_{22}\bigr)\brangle \eta_{a\gamma} \nonumber\\
    & + \tfrac1{32}\blangle \bigl( R^{11}_{11}+R^{22}_{22}-2R^{22}_{11}\bigr)^2\brangle \zeta_\gamma
    -\tfrac12 \blangle \bigl(I^{22}_{11}\bigr)^2\brangle \zeta_{\gamma 1}
    -\tfrac12 \blangle \bigl(I^{23}_{11}\bigr)^2+\bigl(I^{23}_{22}\bigr)^2\brangle \zeta_{a\gamma}\nonumber\\
    &+\tfrac12 \blangle \bigl(R^{23}_{11}\bigr)^2+\bigl(R^{23}_{22}\bigr)^2\brangle \zeta_{a\gamma 1}
    +\tfrac12 \blangle R^{23}_{11}I^{23}_{11} + R^{23}_{22}I^{23}_{22} \brangle \zeta_{a\gamma 2}\,,
  \end{align}
\end{subequations}
whereas for the remaining two  functions defined in reference~\cite{Mirizzi:2009},
$\zeta_\gamma=\langle ({\rho}_{12}+{\rho}_{21})^2 \rangle$
and $\zeta_{a\gamma }=\tfrac12\langle ({\rho}_{13}-{\rho}_{31})^2 \rangle+\tfrac12\langle ({\rho}_{23}-{\rho}_{32})^2
\rangle$, we obtain
\begin{subequations}
  \begin{align}
    \Delta \zeta_\gamma & =
    \tfrac18 \blangle \bigl(R^{11}_{11}-R^{22}_{22}\bigr)^2\brangle R_\gamma
    +\tfrac12 \blangle \mathcal{P}^2_{a\gamma}\brangle R_a
    +\tfrac1{32}\blangle\bigl( R^{11}_{11}+ R^{22}_{22} - 2 R^{22}_{11}\bigr)^2\brangle R_p\nonumber\\
    &-\tfrac12\blangle \mathcal{P}_{a\gamma}\bigl(R^{11}_{11}-R^{22}_{22}\bigr)\brangle \eta_{a\gamma}
    +\blangle \tfrac18 R^{22}_{11}\bigl(R^{11}_{11}+R^{22}_{22}\bigr)
    +\tfrac{3}{32}\bigl[\bigl(R^{11}_{11}\bigr)^2+\bigl(R^{22}_{22}\bigr)^2\nonumber\\
    &+2\bigl(R^{22}_{11}\bigr)^2- 2\bigl(I_{11}^{22}\bigr)^2\bigr]
    + \tfrac38 R^{11}_{11} R^{22}_{22}-1\brangle \zeta_\gamma
    -\tfrac 12 \blangle \bigl(I^{22}_{11}\bigr)^2\brangle \zeta_{\gamma 1}
    -\tfrac12 \blangle \bigl(I^{23}_{11}\bigr)^2 + \bigl(I^{23}_{22}\bigr)^2 \brangle \zeta_{a\gamma}\nonumber\\
    &+\tfrac12 \blangle \bigl( R^{23}_{11}\bigr)^2+\bigl( R^{23}_{22}\bigr)^2\brangle \zeta_{a\gamma 1}
    +\tfrac12 \blangle R^{23}_{11}I^{23}_{11} + R^{23}_{22}I^{23}_{22} \brangle \zeta_{a\gamma 2}\,,\\
    \Delta \zeta_{a\gamma} & = -\tfrac12 \blangle \bigl(I^{23}_{22}\bigr)^2\brangle R_\gamma
    -2 \blangle \bigl( I^{23}_{33}\bigr)^2\brangle R_a
    -\tfrac14 \blangle \bigl(I^{23}_{11}\bigr)^2+\bigl(I^{23}_{22}\bigr)^2\brangle (R_p+\zeta_\gamma)\nonumber\\
    &+2\blangle I^{23}_{22} I^{23}_{33}\brangle \eta_{a\gamma}
    +\tfrac12 \blangle \bigl(R^{23}_{11}\bigr)^2\brangle \zeta_{\gamma 1}
    +\tfrac12\blangle \mathcal{P}^2_{a\gamma} - 2 \mathcal{P}_{a\gamma}  R^{33}_{22} +
    \bigl(R^{33}_{11}\bigr)^2 + \bigl(R^{33}_{22}\bigr)^2-2\brangle \zeta_{a\gamma}\nonumber\\
    &-\tfrac12 \blangle \bigl(I^{33}_{11}\bigr)^2 + \bigl(I^{33}_{22}\bigr)^2\brangle \zeta_{a\gamma 1}
    +\tfrac12 \blangle R^{33}_{11}I^{33}_{11}+I^{33}_{22}\bigl(R^{33}_{22}-\mathcal{P}_{a\gamma}\bigr)\brangle \zeta_{a\gamma 2}\,.
  \end{align}
\end{subequations}
If one keeps terms of at most second order in the components of the dispersive Hamiltonian, as has been done in
reference~\cite{Mirizzi:2009}, then $\zeta_{\gamma 1}$, $\zeta_{a\gamma 1}$ and $\zeta_{a\gamma 2}$ decouple from
the evolution of the other expectation values and the system of equations for the variance reverts to that derived in
reference~\cite{Mirizzi:2009}.

On the other hand, if terms of all order in the dispersion and absorption are kept, then
to obtain a closed set of the evolution equations for the variance one is forced to define three additional expectation values,
$\zeta_{\gamma 1}=\langle ({\rho}_{12}-{\rho}_{21})^2 \rangle$, $\zeta_{a\gamma 1}=\tfrac12\langle ({\rho}_{13}+{\rho}_{31})^2 \rangle+
\tfrac12\langle ({\rho}_{23}+{\rho}_{32})^2 \rangle$ and $\zeta_{a\gamma 2}=\tfrac{i}{2}\langle ({\rho}_{13}+{\rho}_{31})({\rho}_{13}-{\rho}_{31})\rangle+\tfrac{i}{2}\langle ({\rho}_{13}-{\rho}_{31})({\rho}_{13}+{\rho}_{31})\rangle
+\tfrac{i}{2}\langle ({\rho}_{23}+{\rho}_{32})({\rho}_{23}-{\rho}_{32})\rangle+
\tfrac{i}{2}\langle ({\rho}_{23}-{\rho}_{32})({\rho}_{23}+{\rho}_{32})\rangle$. The corresponding evolution equations read
\begin{subequations}
  \begin{align}
    \Delta \zeta_{\gamma 1} & =
    -\tfrac12 \blangle \bigl(I^{22}_{11}\bigr)^2\brangle(R_p+\zeta_\gamma)
    + \blangle \bigl(R^{22}_{11}\bigr)^2-1\brangle \zeta_{\gamma 1}
    +\blangle \bigl(R^{23}_{11}\bigr)^2\brangle \zeta_{a\gamma}
    -\blangle \bigl(I^{23}_{11}\bigr)^2\brangle \zeta_{a\gamma 1}\nonumber\\
    &+\blangle R^{23}_{11}I^{23}_{11}\brangle \zeta_{a\gamma 2}\,, \\
    \Delta \zeta_{a\gamma 1} & =
    \tfrac12\blangle \bigl(R^{23}_{22}\bigr)^2\brangle R_\gamma
    +2 \blangle \bigl(R^{23}_{33}\bigr)^2\brangle R_a +
    \tfrac14 \blangle \bigl(R^{23}_{11}+R^{23}_{22}\bigr)^2\brangle (R_p+\zeta_\gamma)
    + 2 \blangle R^{23}_{22} R^{23}_{33}\brangle \eta_{a\gamma}\nonumber\\
    &-\tfrac12 \blangle \bigl(I^{23}_{11}\bigr)^2\brangle \zeta_{\gamma 1}
    -\tfrac12 \blangle \bigl(I^{33}_{11}\bigr)^2+\bigl(I^{33}_{22}\bigr)^2\brangle \zeta_{a\gamma}
    +\tfrac14\blangle \bigl(R^{33}_{11}\bigr)^2 - \bigl(I^{33}_{11}\bigr)^2 + \bigl(R^{33}_{22}\bigr)^2 - \bigl(I^{33}_{22}\bigr)^2\nonumber\\
    &+ R^{33}_{33}\bigl(R^{11}_{11}+R^{22}_{22}\bigr) + 2\mathcal{P}^2_{a\gamma} +
    4 \mathcal{P}_{a\gamma}  R^{33}_{22} -4 \brangle \zeta_{a\gamma 1}\nonumber\\
    &+\tfrac12\blangle R^{33}_{11}I^{33}_{11} + R^{33}_{22}I^{33}_{22} + \mathcal{P}_{a\gamma} I^{33}_{22}\brangle \zeta_{a\gamma 2}\,,\\
    \Delta \zeta_{a\gamma 2} & =
    -\blangle R^{23}_{22} I^{23}_{22}\brangle R_\gamma
    +4 \blangle R^{23}_{33} I^{23}_{33}\brangle R_a
    -\tfrac12 \blangle R^{23}_{11} I^{23}_{11} + R^{23}_{22} I^{23}_{22}\brangle (R_p +\zeta_\gamma)
    -2\blangle\mathcal{P}_{a\gamma} I^{33}_{22}\brangle \eta_{a\gamma}\nonumber\\
    &-\blangle R^{23}_{11} I^{23}_{11}\brangle \zeta_{\gamma 1}
    -\blangle R^{33}_{11} I^{33}_{11} + I^{33}_{22}\bigl(R^{33}_{22}-\mathcal{P}_{a\gamma}\bigr)\brangle \zeta_{a\gamma}
    -\blangle R^{33}_{11}I^{33}_{11}+I^{33}_{22}\bigl(R^{33}_{22}+\mathcal{P}_{a\gamma}\bigr)\brangle \zeta_{a\gamma 1}\nonumber\\
    &+\tfrac12\blangle \bigl(R^{33}_{11}\bigr)^2 - \bigl(I^{33}_{11}\bigr)^2 +  \bigl(R^{33}_{22}\bigr)^2 - \bigl(I^{33}_{22}\bigr)^2
    -\mathcal{P}^2_{a\gamma} -2\brangle \zeta_{a\gamma 2}\,.
  \end{align}
\end{subequations}


\providecommand{\href}[2]{#2}\begingroup\raggedright
\endgroup

\end{document}